\documentclass[aps,twocolumn]{revtex4}

\usepackage{graphicx}  
\usepackage{subfigure}
\usepackage{multirow}
\usepackage{tabularx}

\usepackage{fancyhdr}
\usepackage{longtable}
\usepackage{parskip}
\usepackage[T1]{fontenc}
\usepackage{dcolumn}   

\usepackage{bm}        
\usepackage{amsfonts}  
\usepackage{amsmath}   
\usepackage{amssymb}   

\newcommand{\pwisein}{\left\{ \begin{array}{ll}}
\newcommand{\pwiseout}{\end{array}\right.}

\newcommand{\SupInf}{Supplementary Section}

\usepackage{siunitx}
\newcommand{\SIadj}[2]{\SI[number-unit-product={\text{-}}]{#1}{#2}}

\usepackage[colorlinks=true, allcolors=blue]{hyperref}

\begin{document}

\title{Scattering invariant modes of light in complex media}

\author{Pritam Pai$^1$}
\author{Jeroen Bosch$^1$}
\author{Matthias K\"uhmayer$^2$}
\author{Stefan Rotter$^2$}
\altaffiliation{stefan.rotter@tuwien.ac.at}
\author{Allard P.\ Mosk$^1$}
\altaffiliation{a.p.mosk@uu.nl}

\affiliation{$^1$Debye Institute for Nanomaterials Science, Utrecht University, 3508 TA Utrecht, the Netherlands\\
$^2$Institute for Theoretical Physics, TU Wien, A--1040 Vienna, Austria
}

\begin{abstract}  

Random scattering of light in disordered media is an intriguing  phenomenon of fundamental relevance to various applications~\cite{Johnson2015}.
While  techniques such as wavefront shaping and transmission matrix measurements~\cite{Mosk2012,Rotter2017} have enabled remarkable progress for advanced imaging concepts~\cite{Yoon2020,Kubby2020,Bertolotti2012,Katz2014_nphot,Kang2017,Jang2018,Horisaki2019,badon2019}, the most successful strategy to obtain clear images through a  disordered medium remains the filtering of ballistic light \cite{Wang2007,Ntziachristos2010,Drexler2015}. 
Ballistic photons with a scattering-free propagation  are, however, exponentially rare and no method so far can increase their proportion.
To address these limitations, 
we introduce and experimentally implement here a new set of optical states that we term Scattering Invariant Modes (SIMs),  
whose transmitted field pattern is the same, irrespective of whether they scatter through a disordered sample or propagate ballistically through a homogeneous medium. 
We observe SIMs that are only weakly attenuated in dense scattering media,
and show in simulations that their correlations with the ballistic light can be used to improve imaging inside scattering materials.

\end{abstract}

\maketitle 

\begin{figure}[!tb]
\centering\includegraphics[width=0.9\columnwidth]{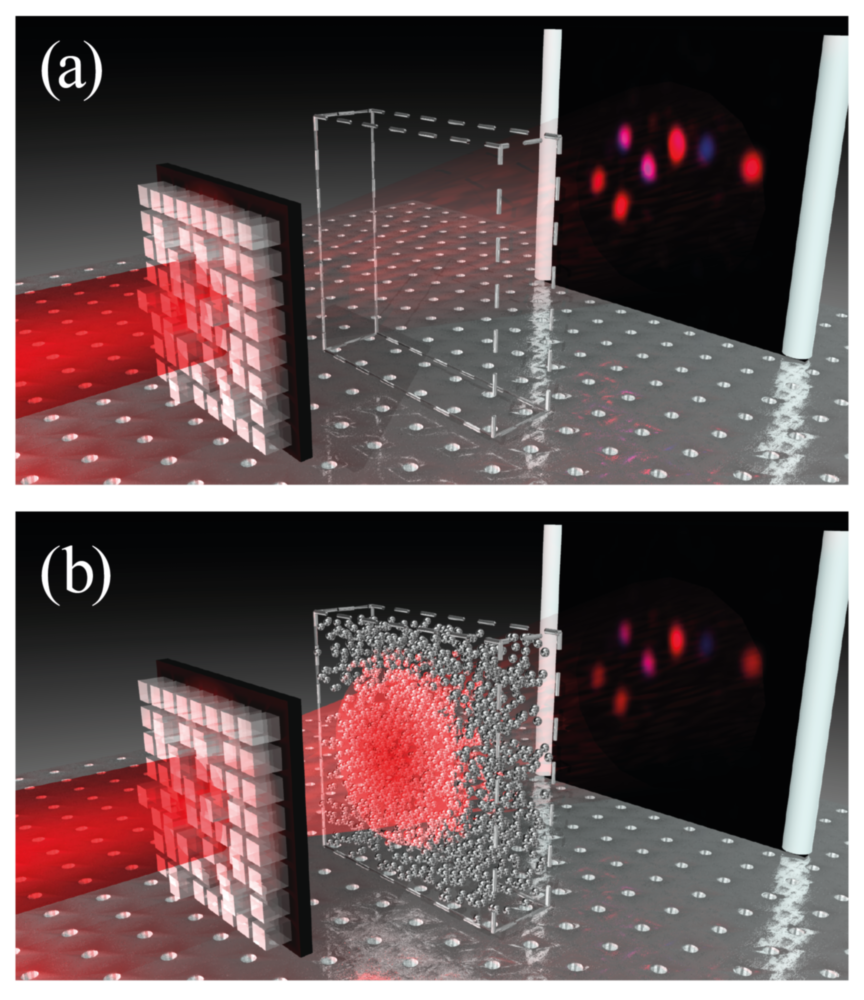}
\caption{\textbf{Illustration of the concept.} A scattering invariant mode (SIM) is generated by a spatial light modulator (SLM) and propagates through \textbf{a},~empty space and \textbf{b},~a scattering sample. The SLM is configured identically in both cases. The SIM is defined so that its transmitted field remains unchanged by the presence of the disordered medium apart from a reduction in overall brightness and a global phase shift, expressed by the complex SIM-eigenvalue $\alpha$.}
\label{fig:GEV_sketch}
\end{figure}

The concept of transmission eigenchannels in scattering media~\cite{Dorokhov1984, beenakker} has greatly expanded both our understanding of light transport~\cite{pendry1990, akkermans_montambaux_2007, Rotter2017, pena2014, Davy13,miller2019} and our ability to engineer the delivery of radiation across disordered materials~\cite{vellekoop2008prl, Mosk2012, Hsu2017, yu2013, kim2012}. While the occurrence and the statistics of these channels has been well described by random matrix theories~\cite{DMPK, beenakker}, the ballistic contributions to the transmission process are much harder to capture by such tools. We introduce here a new set of
Scattering Invariant Modes (SIMs) that embody the defining feature of ballistic light, and which are transmitted through a disordered medium in the same way as through homogeneous space (see Fig.~\ref{fig:GEV_sketch}). To capture this property for constructing SIMs based on experimentally accessible quantities, we employ the corresponding transmission matrices for a scattering medium ($T_{\rm s}$) and for a scattering-free volume of air ($T_{\rm air}$). The input electric fields of SIMs ($\tilde{E}$) are then just determined by the requirement that their output field patterns are the same, irrespective of which one of the two transmission matrices is used to propagate the field from the input to the output. 
Expressing the in- and outgoing fields as vectors of complex coefficients, that represent the horizontal and vertical field components in a suitable basis set, results in the following constitutive relation for SIMs, 
\begin{equation}\label{gen_eig_eq_inv}
    T_{\rm s} \tilde{E}_{\rm n}= \alpha_n T_{\rm air} \tilde{E}_{\rm n}\,.
\end{equation}

\begin{figure*}[!tb]
\centering\includegraphics[width=0.98\textwidth]{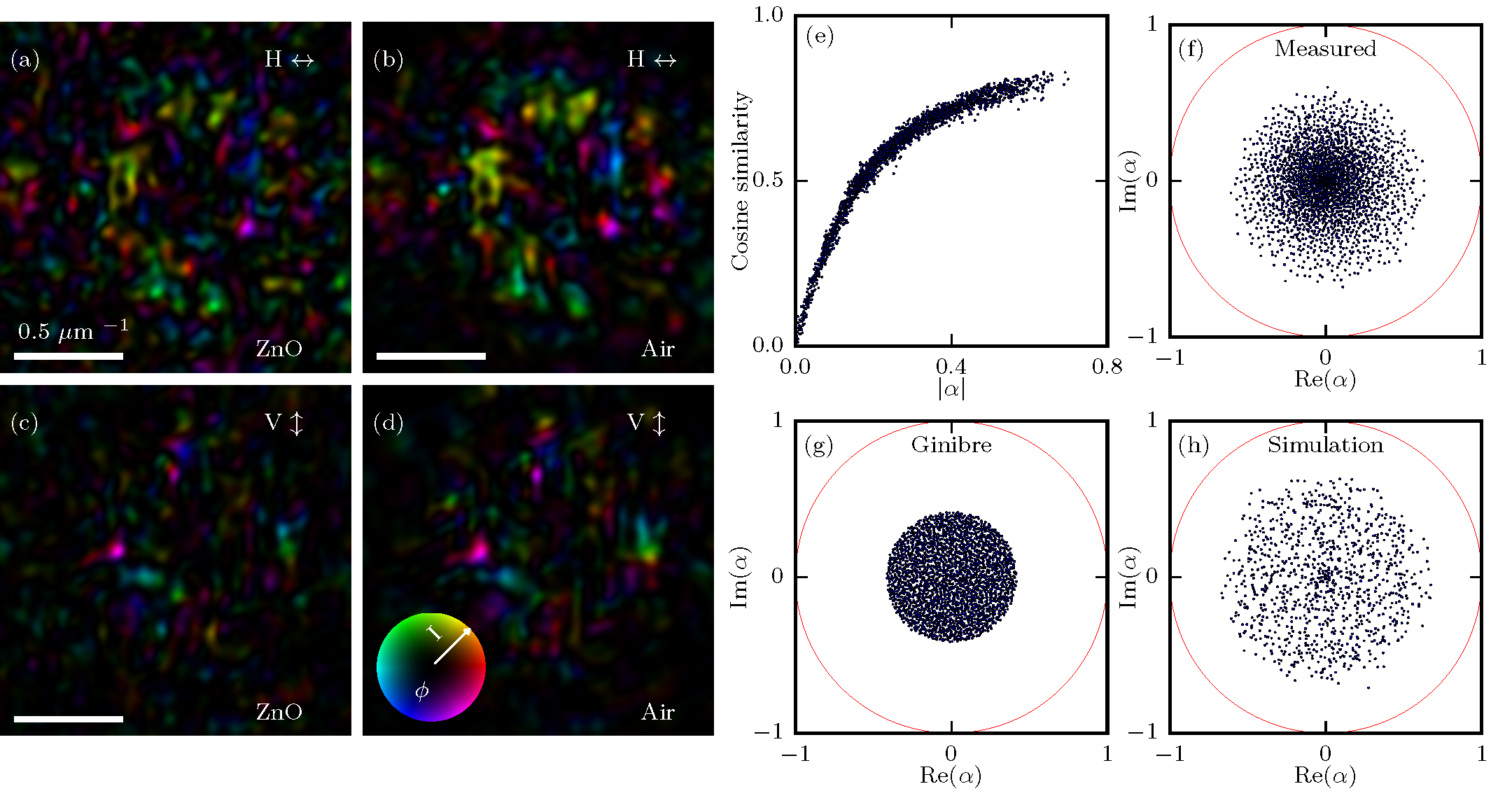}
\caption{\textbf{Experimentally transmitted SIMs and their statistics.} Far field of a SIM transmitted \textbf{a,c}, through a layer of ZnO and \textbf{b,d}, through the same thickness of air in the horizontal (H)  and vertical (V) polarization components, respectively. The color represents phase and the brightness represents amplitude, as specified by the color scheme shown in the inset. The associated cosine similarity of the fields is 0.79 and the SIM eigenvalue is $\lvert \alpha \rvert= 0.64$. \textbf{e},~Cosine similarity of all SIMs propagated through air and through the scattering medium versus the eigenvalue modulus $\lvert \alpha \rvert$. \textbf{f},~Complex generalized eigenvalue spectrum as found from the measured TMs. The unit circle is shown in red. \textbf{g},~Numerically generated distribution of eigenvalues of a Ginibre ensemble. \textbf{h},~SIM eigenvalues from a numerical 2D simulation of a sample with a thickness of 1 mean free path (see Methods).  
}
\label{fig:GEV_proj}
\end{figure*}

\noindent This generalized linear eigenvalue problem demonstrates that SIMs
 emerge as a complete (though not orthogonal) set of input states $\tilde{E}_{\rm n}$. These states are tailored to the scattering medium's transmission matrix  $T_{\rm s}$, and characterized each by a SIM eigenvalue $\alpha_n$. The complex value of $\alpha_n$ 
 quantifies the global amplitude (phase) by which the output field of the corresponding SIM eigenstate $\tilde{E}_{\rm n}$ is attenuated (phase-shifted) when transmitted through the medium rather than through air. To obtain an eigenvalue equation that is numerically stable, we multiply both sides with the Hermitian conjugate of the (in experiments approximately) unitary matrix $T_{\rm air}$, resulting in the SIM-eigenvalue equation
\begin{equation}\label{gen_eig_eq_dag}
   T_{\rm air}^\dagger T_{\rm s} \tilde{E}_{\rm n}= \alpha_n \tilde{E}_{\rm n}\,.
\end{equation}
In this form we see that SIMs are invariant in shape under the operation of forward propagation in the scattering medium followed by back-propagation in air.

To implement this concept in practice, we first measure both transmission matrices $T_{\rm air}$ and $T_{\rm s}$, and insert them in equation~\eqref{gen_eig_eq_dag}. We measure polarization-complete matrices to maximize the level of control over the light field. To generate the resulting coherent fields with controllable amplitude, phase and polarization ellipse on a 2D grid of pixels  we constructed a vector wavefront synthesizer (VWS)~\cite{Bosch2016}. 
A microscope objective projects the pixel array onto the sample surface.
The transmitted light is collected by  a second microscope objective and recorded by a vector wavefront analyzer (VWA), which measures amplitude, phase and polarization ellipse on a similar 2D array of pixels. The measurement and analysis method is reported in detail in Ref.~\cite{Pai20} and \SupInf~1. 

Our scattering sample consists of a layer of zinc oxide (ZnO) nanopowder, deposited on a glass slide, part of which is cleaned to act as a scattering-free reference medium. The scattering layer was inspected to verify the absence of holes. 
The  sample is mounted on a calibrated stage to reversibly exchange the sample for a clean substrate. 

In a first demonstration we use a scattering sample with a thickness of $\SI{1.6\pm0.5}{\micro\meter}$, corresponding to about 2  mean free paths (for these sub-wavelength scatterers  scattering mean free path $\ell$ equals the transport mean free path $\ell_t$). The ballistic transmission is only {12}\%, as found from  the diagonal elements of the transmission matrix. In Fig.~\ref{fig:GEV_proj}(a-d) we show the transmitted field of a SIM with $\lvert \alpha \rvert= 0.64$ projected through both media using the VWS. The measured intensity transmittance of this state is 31\%, which due to imperfect projection is lower than the ideal transmittance $|\alpha|^2=0.41$. 
The fields transmitted through air ($E_{\rm air}$) and the scattering medium ($E_{\rm scat}$) are visually very similar in both polarization components, quantified by a normalized field overlap (cosine similarity) of $|E_{\rm scat}^* \cdot E_{\rm air}|/(|E_{\rm scat}||E_{\rm air}|)$ = 0.79.
The measured cosine similarity for all the projected SIMs, displayed in Fig.~\ref{fig:GEV_proj}(e), shows a strong dependence on $\lvert \alpha \rvert$. We find that the similarity increases monotonically with $\lvert \alpha \rvert$ to reach values up to about 0.82.
Ideally one expects the cosine similarity between SIMs projected through the different media to be unity. In the experiment, the effects of noise,  imperfect projection and the nonunitarity of the experimental $T_{\rm air}$ reduce this value, especially for low $|\alpha|$.

The complex SIM-eigenvalue spectrum is shown in Fig.~\ref{fig:GEV_proj}(f). The phase of $\alpha$ in this diffuse spectrum is distributed approximately isotropically. The density of eigenvalues gradually decays with distance from the origin until around $\lvert \alpha \rvert = 0.66$, corresponding to 3.5 times the ordinary ballistic transmission.
For comparison, we show in Fig.~\ref{fig:GEV_proj}(g) the eigenvalue spectrum of a complex Gaussian random matrix with the same average transmission, which is homogeneous inside a sharply bounded Ginibre disk. 
Further statistics of the measured transmission matrix can be found in \SupInf~2.

In Fig.~\ref{fig:GEV_proj}(h) we show the complex eigenvalue spectrum of the numerically calculated matrix $T_{\rm air}^\dag T_{\rm s}$ of a sample of one mean free path in thickness, which includes the effect of the finite NA of the objectives~\cite{stone2013}. We observe a nonuniform distribution similar to the experimental data.
The deviation of both the experimental and the simulated data from the eigenvalue spectrum of a random matrix shows that the eigenvalues reflect the sample-specific details of the scattering process. The occurrence of relatively large SIM-eigenvalues, which are exponentially rare in random matrix theory~\cite{Forrester2007_prl}, is particularly noteworthy since the corresponding SIMs feature output speckle patterns with the largest similarity values and with the highest transmission through the scattering sample.

In many cases one is only interested in the field transmitted into a few modes, such as pixels making up a sparse image.
We generate the corresponding sparse SIMs by numerically back-propagating the sparse target field through the sample (by applying the experimentally measured $T_{\rm s}^\dag$), and separately through the reference (by applying the experimentally measured $T_{\rm air}^\dag$), and superposing the two (typically completely uncorrelated)  incident fields thus obtained (see Methods).

\begin{figure}[tb]
\centerline{\includegraphics[width=0.98\columnwidth]{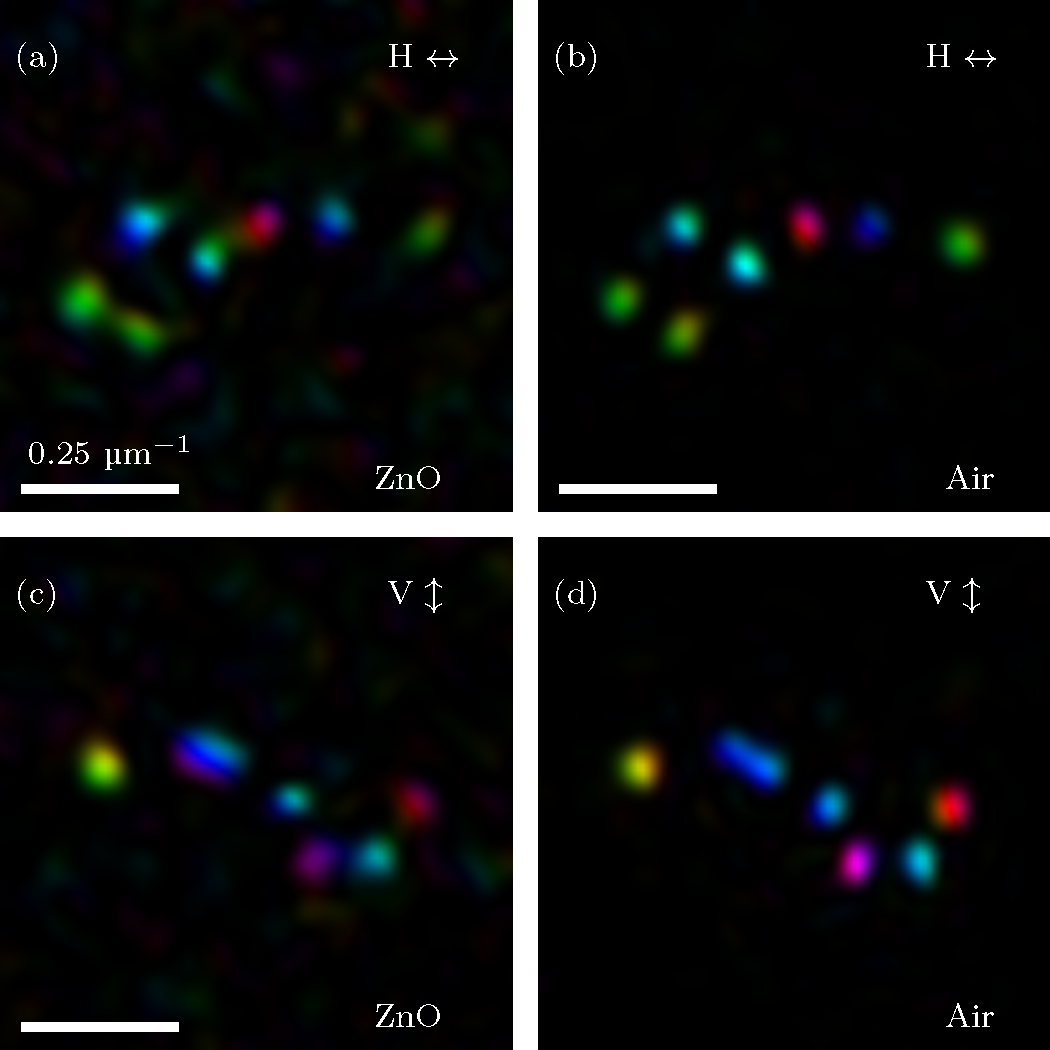}}
\caption{\textbf{Experimental realization of a sparse SIM.} Far field intensity of the horizontal (H) and vertical (V) polarization components of a  sparse SIM transmitted through \textbf{a,c},~a $\SIadj{5.2\pm0.5}{\micro\meter}$-thick ZnO powder, and  \textbf{b,d},~through the same thickness of air,
shaped to resemble the stellar constellations of Ursa Minor (top row, H polarization) and Ursa Major (bottom row, V polarization).
The color scheme is the same as in Fig.~\ref{fig:GEV_proj}. 
}
\label{fig:sparseGEV}
\end{figure}

In Fig.~\ref{fig:sparseGEV} we show an example of a sparse SIM propagated through a ZnO powder sample and through the same thickness of air. 
The SIM shown here is constructed to display a point pattern where  we independently control the amplitude, phase and polarization in each spot. While a low-intensity uncorrelated speckle is visible in the background, the similarity between the high-intensity spots is striking. 
This principle is easily generalized to propagate images with controlled phase and polarization through two different complex media using only one incident field. 

So far several transmission matrix-based methods have proven valuable for imaging through scattering media.  The intriguing new possibility 
 is that SIMs allow to improve imaging  \textit{inside} the complex medium. Externally, each SIM appears to propagate through a semi-transparent medium with transmittance $|\alpha|^2$ and a phase of $\arg(\alpha)$ relative to propagation through air.
For sparse scattering media that have scatter-free areas, we show in \SupInf~3 that SIMs with high values of $|\alpha|$ preferentially propagate through the scatter-free areas, just like ballistic light would do inside the medium.

Inside dense scattering media without holes or gaps, we find, using both 2D and 3D wave simulations, that SIMs remain correlated with the ballistic component of the incident light up to a depth $z$ of several mean free paths, but are gradually shifted in phase and amplitude as they propagate . In \SupInf~4 we show numerically that the phase shift they accrue while propagating to the depth $z$ inside the medium of thickness $L$ turns out to be well described by $\arg(\alpha)\times z/L$. By compensating this phase  shift we synthesize a ``SIM-corrected'' wave that focuses better than an uncorrected wavefront deep inside the scattering sample. Specifically, to produce a focus at depth $z$ inside the sample, we first back-propagate an ideal focus to the input surface using the known air transmission matrix $T_{\rm air}(z)$ from the surface to the scattering layer at depth $z$. We then decompose the back-propagated focus field into SIMs, which are then individually corrected by the phase shift expected for their specific values of $\alpha$ and $z$. An  amplitude correction  provides more weight to those SIMs with small phases of $\alpha$ since they feature a higher fidelity with respect to the field in the reference medium. These adjusted SIM amplitudes then enter the corrected incident field (see \SupInf~4 and 5 for full details) and are numerically propagated through the scattering medium to check their focusing performance with respect to the uncorrected field. Importantly, our correction procedure solely relies on data from the experimentally accessible external transmission matrix \cite{popoff2010}.

\begin{figure}[!tb]
\centering\includegraphics[width=\columnwidth]{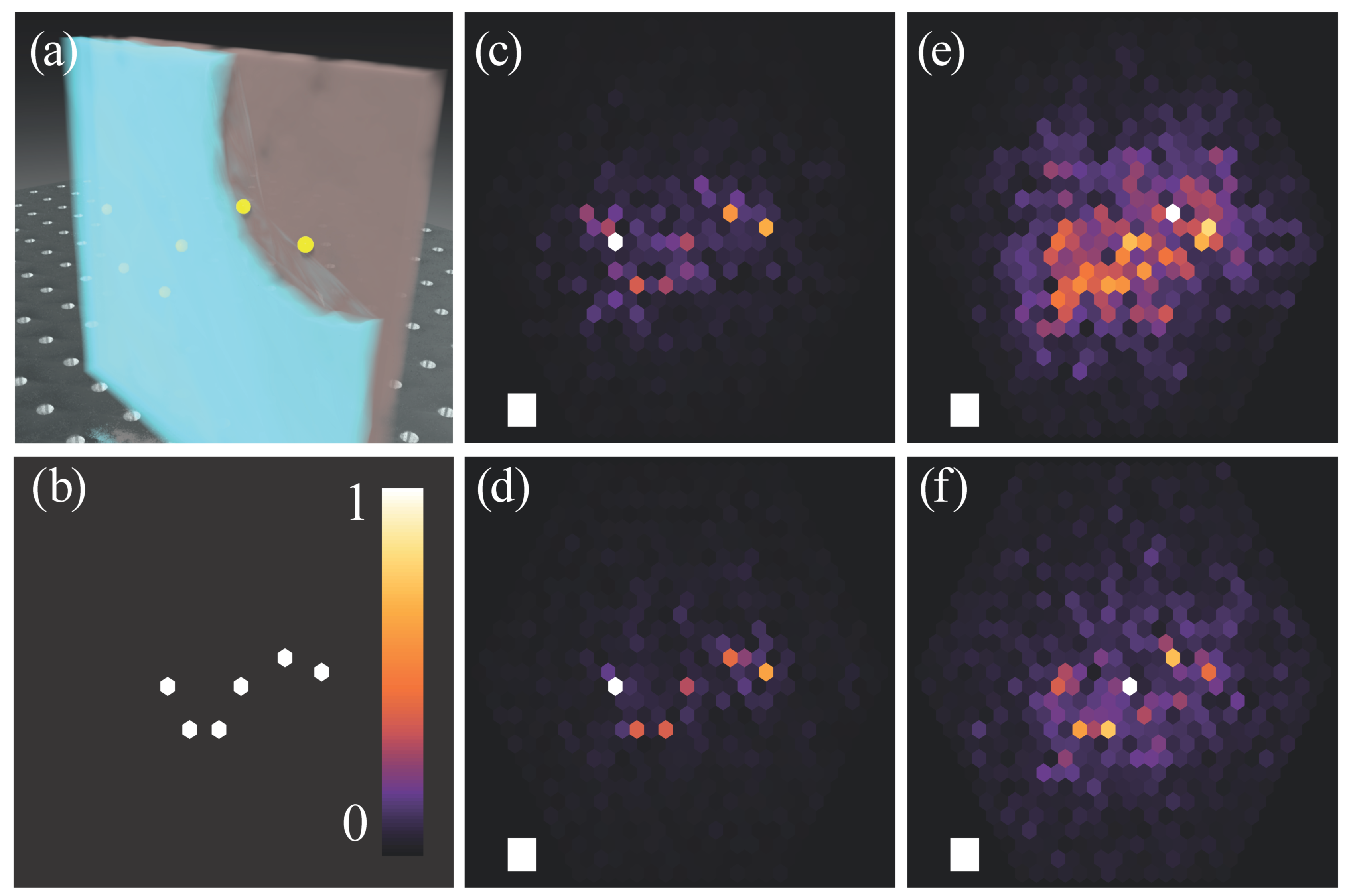}
\caption{\textbf{Simulations of imaging using SIMs.}
\textbf{a},~Sketch of the simulation setup: a fluorescent object is sandwiched between two scattering layers (one peeled open for clarity), each of thickness $L/2$. \textbf{b},~The simulated object consists of six point-like particles. \textbf{c},~Its uncorrected, and, \textbf{d},~its corrected image using SIMs for a system thickness of $L=5.4\, \ell$. \textbf{e,f},~The same plots for $L=7.1\, \ell$. The squares indicate the size of the wavelength in the simulation.}
\label{fig:Imaging}
\end{figure}

In Fig.~\ref{fig:Imaging} we show typical simulation results for scanning excitation imaging of a fluorescent object, sandwiched between two scattering layers as depicted in Fig.~\ref{fig:Imaging}(a,b). The layers are strongly and isotropically scattering, with a scattering mean free path $\ell$ of about 4 wavelengths.
In the most optical thin case considered here the uncorrected image in Fig.~\ref{fig:Imaging}(c) is already heavily distorted by speckle. 

In Fig.~\ref{fig:Imaging}(d) we demonstrate the effect that the SIM-based correction of the excitation beam has using the procedure described above. Panels (e,f) show the corresponding images for a more strongly scattering system. The improvement in imaging is quantified by the correlation coefficient between the recovered image and the true object, averaged over many realizations as detailed in Table~\ref{table:statisticssmall}. We see that a significant improvement in image quality is possible up to a large thickness of $L=10 \ell$. SIM based correction of the incident wave may be combined with confocal detection or multiphoton methods to allow even deeper imaging.

\begin{table}[h!]
\centering
\begin{tabular}{|c|c|c|c|}
 \hline
Thickness $L/\ell$& $C_{\rm uncorr.}$ (sd) & $C_{\rm SIM}$ (sd) & $N$ \\
 \hline
 5.4 & 0.728(0.03) & 0.802(0.03) & 23 \\
\hline
 7.1 & 0.500(0.07) & 0.606(0.07) & 50 \\
\hline
 9.0 & 0.326  (0.07) & 0.395 (0.09) & 52   \\
\hline
\end{tabular}
\caption{\textbf{Quantifying the use of  SIMs for deep imaging.} Mean Pearson's correlation coefficient (standard deviation in parentheses) between images (uncorrected and SIM-corrected) and the true object, averaged over many different disorder configurations $N$ (see rightmost column). \label{table:statisticssmall}}
\end{table}


In summary, we introduce  the concept of scattering invariant modes (SIMs), which produce the same transmitted field profile through a multiple scattering sample as when propagating through a reference medium. We successfully generate SIMs experimentally and find a high  similarity between the fields propagated through a layer of ZnO powder and through the same thickness of air. We show explicitly how to use this feature to achieve free-space transmission of simple images across strongly disordered media. 
Moreover, SIMs remain correlated with the ballistic light inside a medium and can be used to improve imaging quality under difficult conditions. These remarkable properties make SIMs an attractive new set of tools both from a fundamental perspective as well as for applications in imaging through complex media. 

\section*{Acknowledgements}
{We acknowledge fruitful discussions with P.\ Ambichl, D. Bouchet, S. Faez, F.\ Salihbegovic, and S.\ Steinhauer during early stages of this project. S.R.\ wishes to thank the Austrian Science Fund (FWF) for support under project WAVELAND (Grant No.\ P32300). This project was supported by NWO-Vici grant 68047618. Experiments were supported by P. Jurrius, D. Killian and C.R. de Kok. The computational results presented in this paper were achieved using the Vienna Scientific Cluster (VSC). }

\section*{Author contributions}
The experiments were designed by A.P.M., P.P. and J.B  and implemented by {P.P. and J.B.} The 2D full wave numerical simulations and the theoretical analysis were carried out by M.K. and the 3D calculations by A.P.M. S.R.~proposed the idea and supervised the theoretical research. All authors analyzed the results and contributed to the writing of the manuscript.



\section*{Methods}
\textbf{Vector wavefront analyzer}. The transmitted field is measured with angle-offset holography~\cite{Cuche00}, for both polarization components separately. The speckle field transmitted through the scattering sample is collected with a high-numerical aperture immersion objective (NA=1.4). The two polarization components are imaged on separate cameras, and interfered with a reference beam that is incident at a small angle. The local amplitude and phase is calculated from the interference pattern.

The TM measurements are carried out by sending incident waves with 1141 different wavevectors, on a hexagonal grid in $k$-space, for two orthogonal polarizations. The transmitted amplitudes are resampled to the same basis set yielding matrices of dimension $2282 \times 2282$. Both the incident and the transmitted fields are sampled on a hexagonal lattice in Fourier space with a lattice constant chosen to minimize overlap between adjacent lattice points.

\medskip
\noindent \textbf{Construction of sparse SIMs}.
Sparse SIMs emerge from the medium as a sparse field coupling to a relatively small number   $M$  modes of interest. For simplicity we assume these modes correspond to waves $E_i$ focused in target points $\{y_i\}$. The basis of incident waves is chosen to be  $\{u_i,v_i\}$, with $u_i =T_{\rm air}^\dagger E_i /|T_{\rm air}^\dagger E_i|$ and $v_i =T_{\rm s}^\dagger E_i/|T_{\rm s}^\dagger E_i|$, i.e., this basis contains the waves that are optimally focused on the target points~\cite{vellekoop2008prl}, either through the air or through the sample. Assuming $M \ll N$ and that the target modes are non-overlapping, the dot product between the $u_i$ and $v_i$ can be neglected and we have 
$T_{\rm s}v_i \approx \bar{T} E_i$, where $\bar{T}$ is the average transmission of the scattering sample~\cite{vellekoop2008prl}. Similarly, $T_{\rm air}u_i \approx E_i$ since the air has transmission unity.  
Under these assumptions an approximate SIM can be constructed as 
$E_{\rm in}(\alpha)=  C\sum_i   (u_i +\alpha v_i/\bar{T})$.
Here, $C$ is a normalization constant and $\alpha$ is the desired generalized eigenvalue. The procedure is equivalent to taking a weighted sum of a field shaped to project the image through air and a field shaped to project the same image through the scattering sample.

\medskip
\noindent \textbf{Numerical simulations}. We numerically solve the two-dimensional scalar Helmholtz equation $[\Delta + n^2(\vec{r}) k_0^2] \psi (\vec{r}) = 0$ using the finite element method \cite{Schoeberl2014} (\href{https://ngsolve.org}{https://ngsolve.org}), where $\Delta$ is the Laplacian in two dimensions, $n(\vec{r})$ is the refractive index distribution, $\vec{r} = (x,y)$ is the position vector, $k_0 = 2\pi/\lambda$ is the free space wavenumber with $\lambda$ being the wavelength and $\psi (\vec{r})$ is the $z$-component of the TE-polarized electric field. 

To model the experimental systems we use a scattering region which is longitudinally and transversally attached to leads featuring hard-wall boundary conditions in order to use a waveguide mode basis, where we use $k_0 = 2\pi / \lambda = 1000.5 \pi /W$ with $W$ being the width of the longitudinal input and output lead. Perfectly matched layers (PMLs) are then added to these open ends in order to absorb the outgoing waves without any backreflections, thus mimicking semi-infinite leads. Since the scattering matrix is only calculated between the input and output port, the remaining PMLs at the top and bottom leads of our geometry serve as loss channels arising in the experiment due to a limited numerical aperture. The length of the scattering region $L$ is then adjusted according to the wavelength in the experiment, i.e., we use the same ratio of $L/\lambda$ and  as in the experiment. The ZnO nanoparticles are modeled by circular scatterers with a refractive index of $n_\text{scat} = 2$, where their diameter is again of the same ratio $d_\text{scat}/\lambda$ as in the experiment. Since these ZnO nanoparticles tend to stick together, we mimic this behavior by using circular scatterers of larger sizes.

The densely filled samples in Fig.~\ref{fig:GEV_proj} are simulated with a scattering region of length $L = \lambda (L^\text{exp}/\lambda^\text{exp})$ with $L^\text{exp}$ = \SI{2.1}{\micro\meter} which is 40\% filled with circular scatterers whose diameter are $d_\text{scat} = n \lambda (d_\text{scat}^\text{exp}/\lambda^\text{exp})$. Here, $d_\text{scat}^\text{exp} \approx $\SI{200}{\nano\meter}, $\lambda^\text{exp} = $ \SI{633}{\nano\meter} and $n = 1, 2$, where we choose that every scatterer size fills out the same fraction of the total area of the scattering region. To simulate free space, we remove all the scatterers which leaves us with an empty scattering region in which waves can escape through the transversally added leads featuring PMLs which absorb the outgoing waves. 

Last, we use mesoscopic transport theory to calculate the scattering matrices of these systems between the input and the output lead which are then used to calculate the scattering invariant modes from equation~\eqref{gen_eig_eq_dag}.

\newpage

\section*{Appendices}  

In this supplement we present additional details on the experimental procedure and setup (Section~1), the statistics of the SIM eigenvalues and singular values of the transmission matrices (Section~2), and the property of SIMs to avoid scattering in very sparse samples (Section~3).

In Section~4 we describe our method for 3D simulations of transmission matrices, which we use in Section~4.2 to study the fidelity of SIMs inside the medium, and in Section~4.3 to simulate deep imaging using SIMs. Finally in Section~5 we show 2D full-wave simulations that corroborate the 3D results.


\begin{figure}[b!]
	\centerline{\includegraphics[width=0.98\columnwidth]{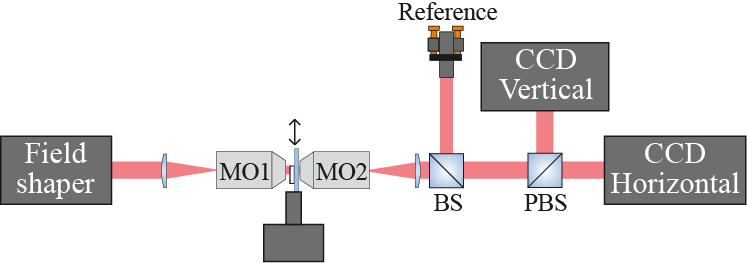}}
	\caption{Experimental setup. A hologram is generated on the Field shaper, and projected on the surface of the sample with a lens and microscope objective (MO1). The sample is mounted on a automated stage, which allows controlled motion of the sample. The transmitted light is collected through a second microscope objective (MO2) and imaged on two cameras (CCD Vertical and CCD Horizontal). A diagonally polarized reference beam is overlapped with the light transmitted from the sample with a beamsplitter (BS). We split the polarization components with a polarizing beamsplitter (PBS) between the two cameras.}
    \label{fig:Setup_Schematic}
\end{figure}

\begin{figure}[tb]
	\centerline{\includegraphics[width=0.98\columnwidth]{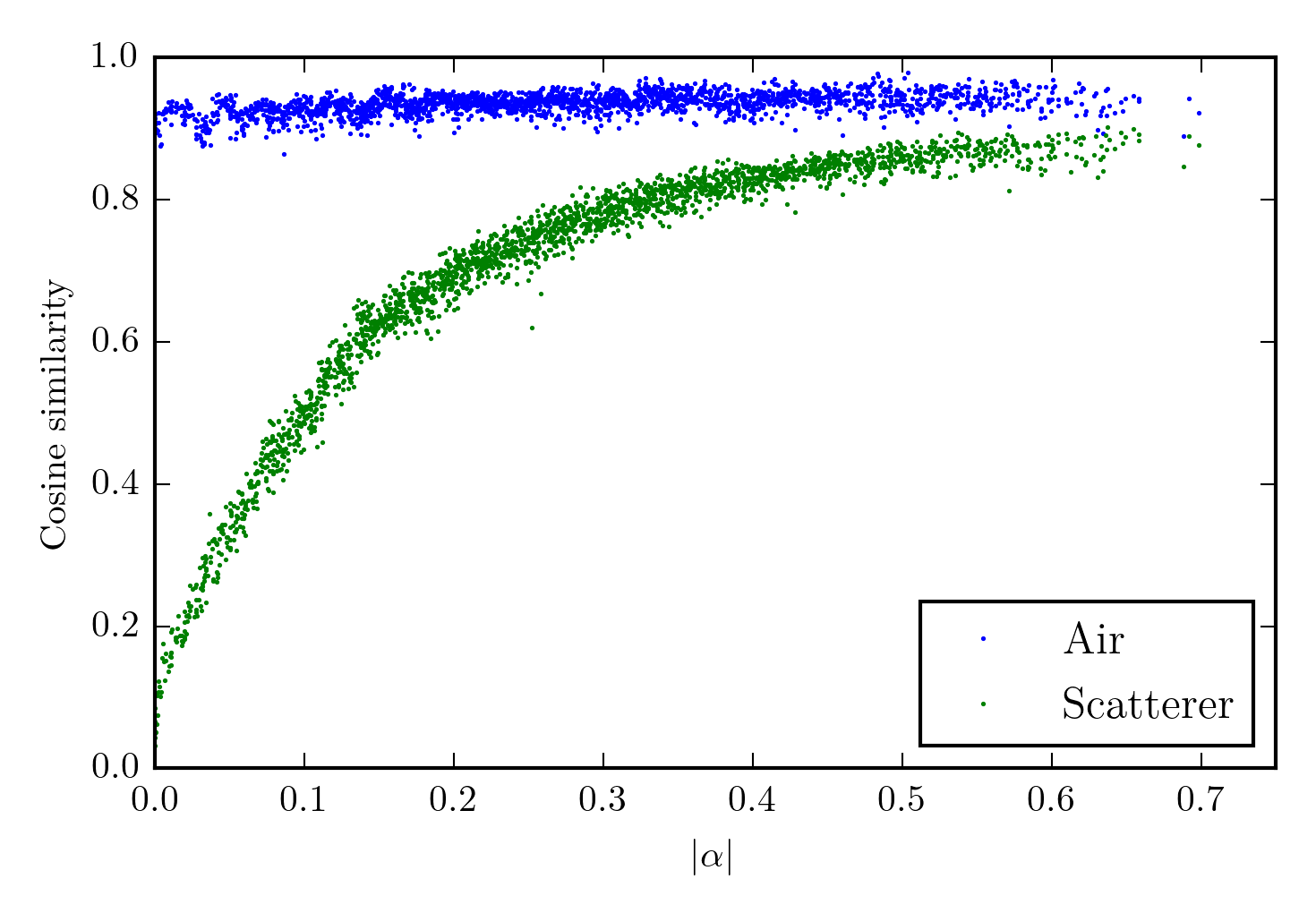}}
	\caption{Cosine similarity between the physically propagated SIM, and the propagated SIM as predicted by the transmission matrix, for both media.}
    \label{fig:sup:cosine}
\end{figure}

\begin{figure*}[tb]
	\centerline{\includegraphics[width=0.75\textwidth]{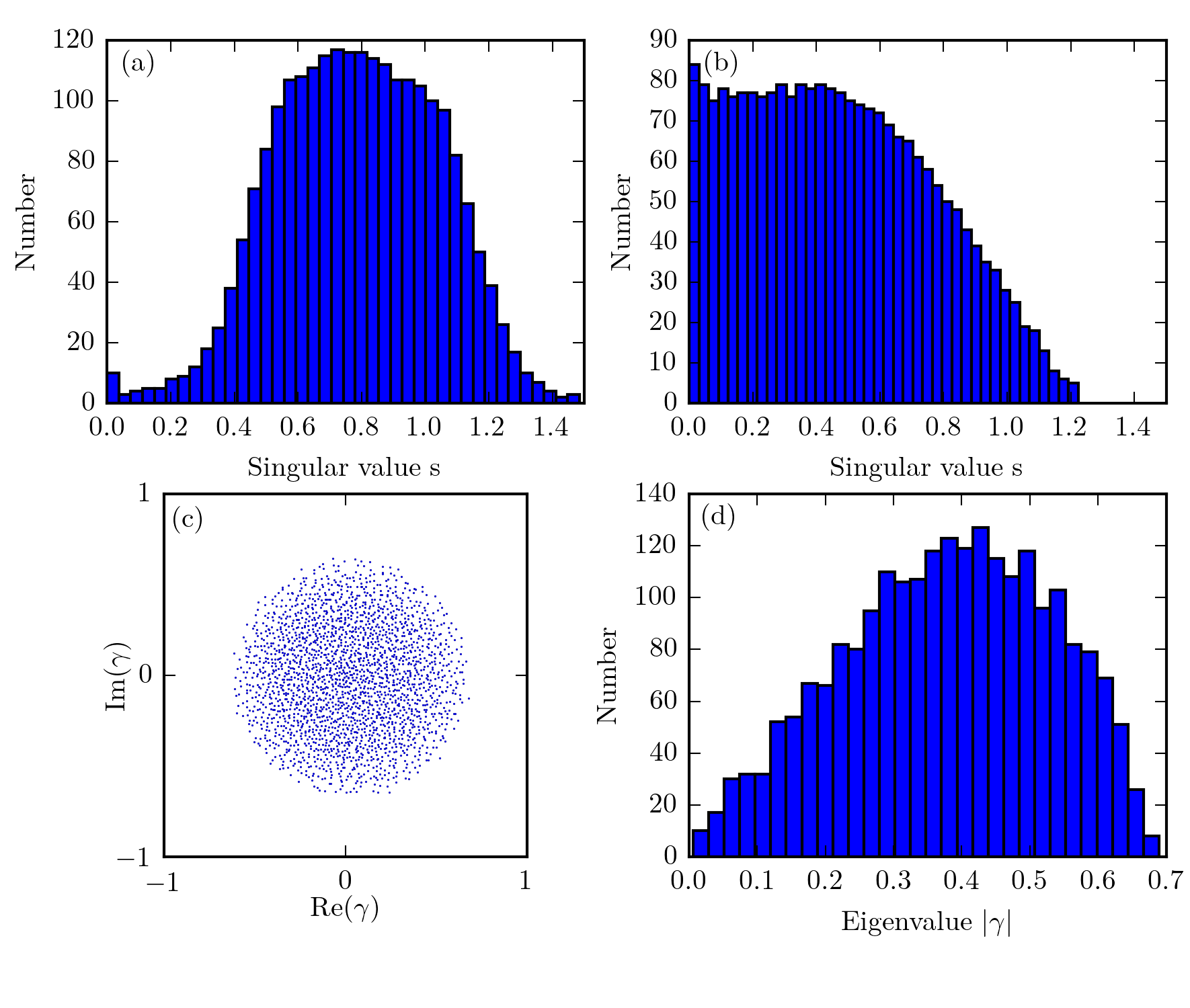}}
	\caption{Singular value histograms for (a)~a zero-thickness air measurement and (b)~for the scattering sample used in  Fig.~2. in the main text. (c) Complex eigenvalue distribution of the transmission matrix, and (d) the corresponding histogram of the eigenvalue amplitudes.}
    \label{fig:Ssup:singularair}
\end{figure*}

\section{Experimental details and setup}

We show the experimental apparatus, used to measure the transmission matrices and project the SIMs through the sample in Fig.~\ref{fig:Setup_Schematic}. Light is generated by a external-cavity diode laser (TLB-6712) at \SI{771}{\nano\meter} and coupled into single mode fibers. We generate a hologram on the Field shaper with a DMD (ViALUX V-9600) and Lee holography~\cite{Lee1974}. We control the horizontal and vertical polarization component simultaneously, and control both amplitude and phase of the hologram. The DMD is imaged on the sample through a high-NA objective (MO1, NA 0.95, Zeiss N-Achroplan) and a 750mm lens. The sample is deposited on a coverslide. The coverslide with the sample is mounted on an automated 2D-stage (Smaract), which allows us to move with nanometer precision. This allows us to move between a clear and cleaned area on the cover slide and a scattering area. 
\begin{figure}[tb]
	\centerline{\includegraphics[width=0.98\columnwidth]{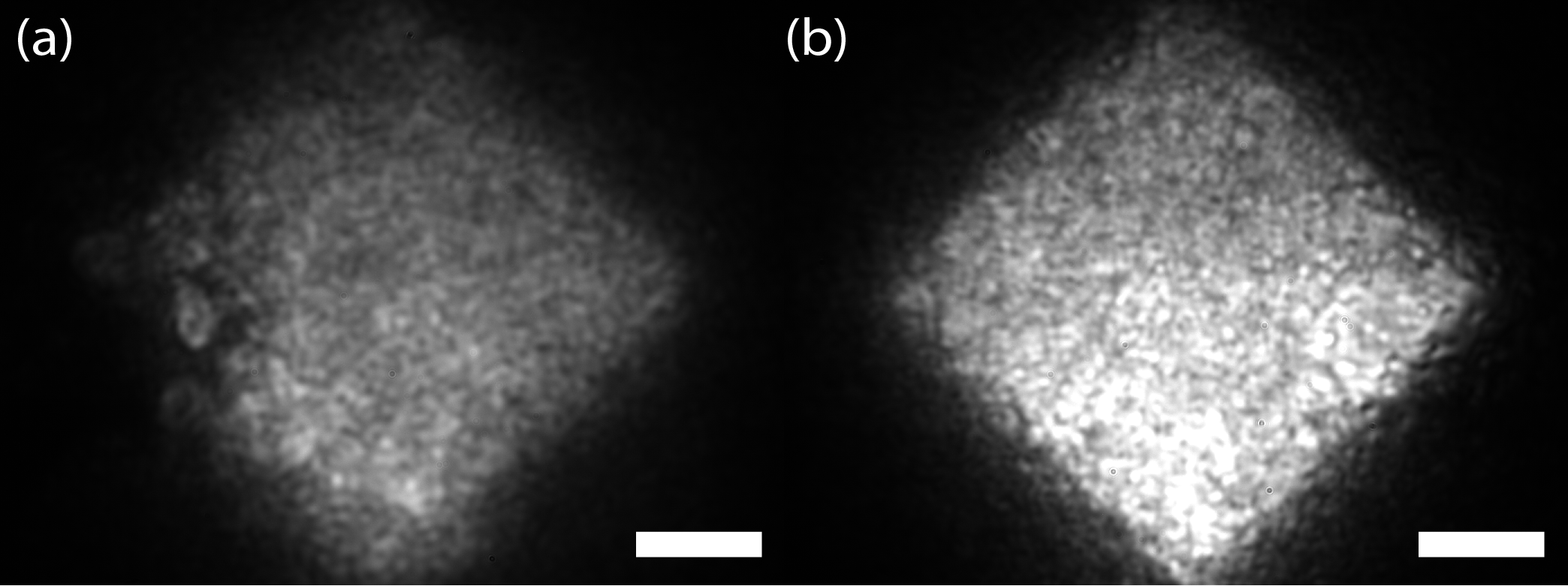}}
	\caption{White light images of the scattering medium in (a)~transmission and (b)~reflection. A square shaped illumination pattern is used which shows no visible holes, or large inhomogeneities of the sample thickness. Scale bar, \SI{5}{\micro\meter}.}
    \label{fig:Ssup:sample_surface_thin}
\end{figure}
The transmitted light is collected through a high-NA immersion objective (MO2, NA1.4, Zeiss Plan-Apochromat). The objectives are mounted on calibrated closed-loop piezo stages, which allows us to refocus the objectives and measure the sample thickness in-situ. The backside of the sample is imaged onto two CCD cameras with MO2, a 750mm lens, and a polarizing beamsplitter (PBS) which allows us to measure both transmitted polarization components simultaneously. A reference beam is combined with the signal beam from the sample through a beamsplitter (BS) which allows us to measure the local amplitude and phase of the transmitted field through angle-offset holography~\cite{Leith62,Takeda82}.

\begin{figure*}[tb]
	\centerline{\includegraphics[width=0.75\textwidth]{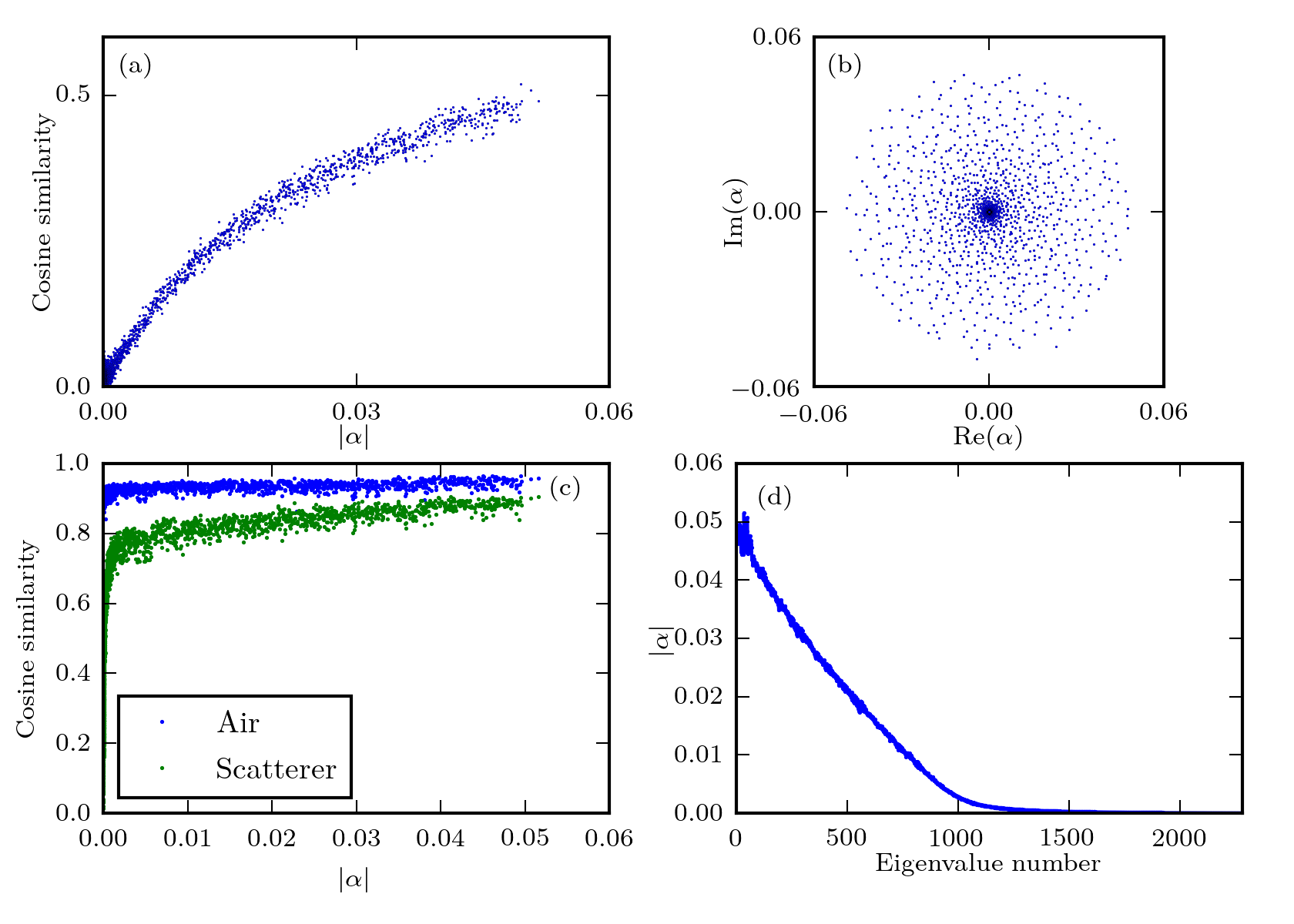}}
	\caption{SIM data for a thick (\SI{21.8}{\micro\meter}) sample. Cosine similarity in (a), the eigenvalues in (b) and the cosine similarity between the propagated field and the projected field in (c). Finally in (d) we plot the absolute value of the eigenvalue versus the eigenvector number.}
    \label{fig:Ssup:thick_SIM}
\end{figure*}
We sample the transmission matrix by projecting circular aperture planar waves with a diameter of \SI{23}{\micro\meter}, of which we vary the incident angle ($k$-vector). We sample the incident $k$-vectors on a hexagonal grid, to increase the sampling density~\cite{Pai20}. The step size between the $k$-vectors is chosen as the minimum distance for which sampled fields with adjacent $k$-vectors do not show a field overlap. The transmitted field is also sampled in $k$-space, on a grid matched to the incident $k$-vectors. The highest sampled $k$-vector corresponds to a NA of approximately 0.9.

To confirm the quality of our wavefront synthesizer we show in  Fig.~\ref{fig:sup:cosine}  the cosine similarity between the field experimentally  propagated  through the scattering medium and the field numerically propagated through the transmission matrix, versus the eigenvalue $|\alpha|$. We see that this similarity is always high, ranging from 0.86 to 0.98, for states projected through air, while it increases with $|\alpha|$ for states projected through the scattering medium. This behavior is explained by the fact that the vector wavefront synthesizer (VWS) is emits a few percent of its output in undesired modes. For low $|\alpha|$ the transmission of the intended component is suppressed by a factor $|\alpha|^2 \ll 1$ while the power in the undesired modes propagate with (on average) the mean diffusive transmission of the sample. As a result the latter dominate the transmitted light.

\section{Distributions of the eigenvalues, SIM-eigenvalues and singular values of the measured transmission matrices}

In Fig.~\ref{fig:Ssup:singularair} we show the statistics of the transmission matrices of a zero-thickness air reference and of the scattering medium corresponding to Fig.~2 in the main text of the manuscript. Fig.~\ref{fig:Ssup:singularair}(a) shows the singular value distribution of the TM of a zero-thickness reference medium, which entails a system where the illumination and detection planes are identical. The transmission singular values are  normalized to the peak column intensity of the TM. We observe a single broad peak centered on 0.8 with a large width of 0.6.  Ideally, for such a transparent system, we  expect a unitary transmission matrix with all singular values equal to 1. However, the peak broadens due to  noise and slight nonuniformity of the illumination. The pedestal is dependent on the sampling criterion of the incident modes~\cite{Pai20}. This explains the occurrence of singular values above 1 in the histograms.

In Fig.~\ref{fig:Ssup:singularair}(b) we show the singular value histogram of the scattering sample. The occurrence of very high singular values indicates that the sample is only a few mean free paths in thickness. The white light images of the reflected and transmitted light indicate there are no physical holes present in the sample, as seen in Fig.~\ref{fig:Ssup:sample_surface_thin}.

The standard eigenvalues of the TM of the scattering medium (obtained by eigenvalue decomposition of $T_s$ without using the air matrix) are shown in the complex plane  and in a histogram of the modulus in Fig.~\ref{fig:Ssup:singularair}(c) and (d), respectively. Comparing Fig.~\ref{fig:Ssup:singularair}(b) to Fig.~2 in the main text it is obvious that the standard eigenvalues are more evenly distributed (i.e., more like those of a random matrix from the Ginibre ensemble) than the SIM eigenvalues.

\subsection{Thick scattering samples}
\label{sec:Ssup:thick_sample}
We have also performed SIM measurements on much thicker samples. In Fig.~\ref{fig:Ssup:thick_SIM} we show the results for a  sample with a thickness of \SI{21.8}{\micro\meter}, or about 30 transport mean free paths. We find that approximately half of the eigenvalues are close to zero, and the other eigenvalues are spread on the complex plane like in the thin scattering sample. We find a high cosine similarity of up to 0.5, even for SIMs with a very low eigenvalue of below 0.05. Even for these very thick samples we find very high cosine similarities between the physically propagated field, and the field as predicted by the transmission matrix.

\begin{figure*}[tb]
	\centerline{\includegraphics[width=0.98\textwidth]{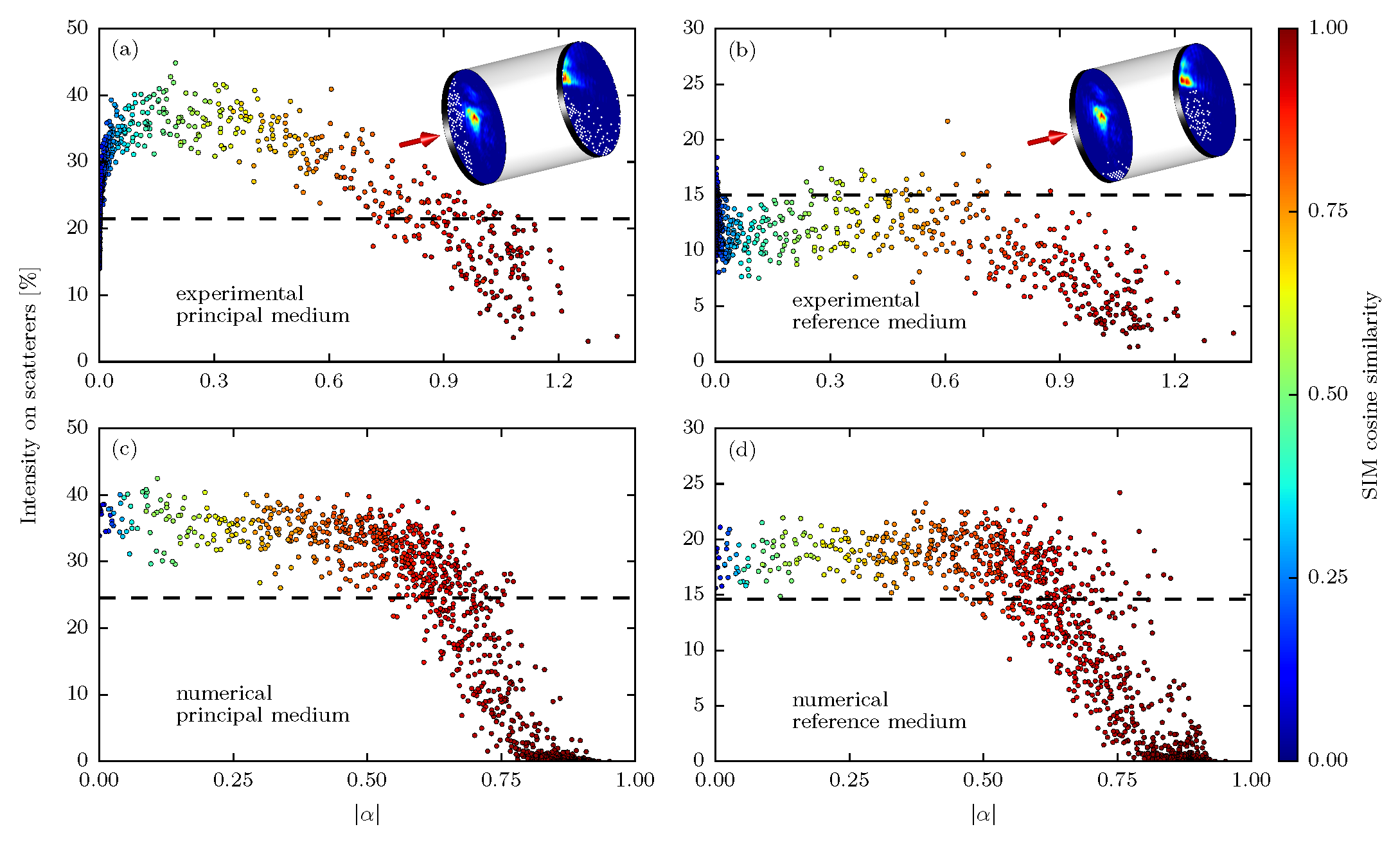}}
	\caption{SIMs in sparse samples. (a,b), Overlap of the SIM intensity with the scatterers versus SIM eigenvalue magnitude (a) on the principal and (b) on the reference medium. The color of the dots represents the cosine similarity of the transmitted fields through both media. The dashed horizontal line represents the average surface coverage. The insets show the measured intensity distributions of a SIM featuring a high value of $|\alpha|$ on both layers of the corresponding sample, where the color corresponds to the normalized intensity on each layer and the white dots mark the scatterer positions. (c,d), Corresponding results obtained from simulations with similar surface coverage. Both the experimental and the numerical results confirm that the highest values of $|\alpha|$ are not only associated with a very high degree of similarity, but also with a low overlap of the corresponding field intensity with the scatterers. 
	}
    \label{fig:GEV_avoidance}
\end{figure*}
\begin{figure}[tb]
	\centerline{\includegraphics[width=0.98\columnwidth]{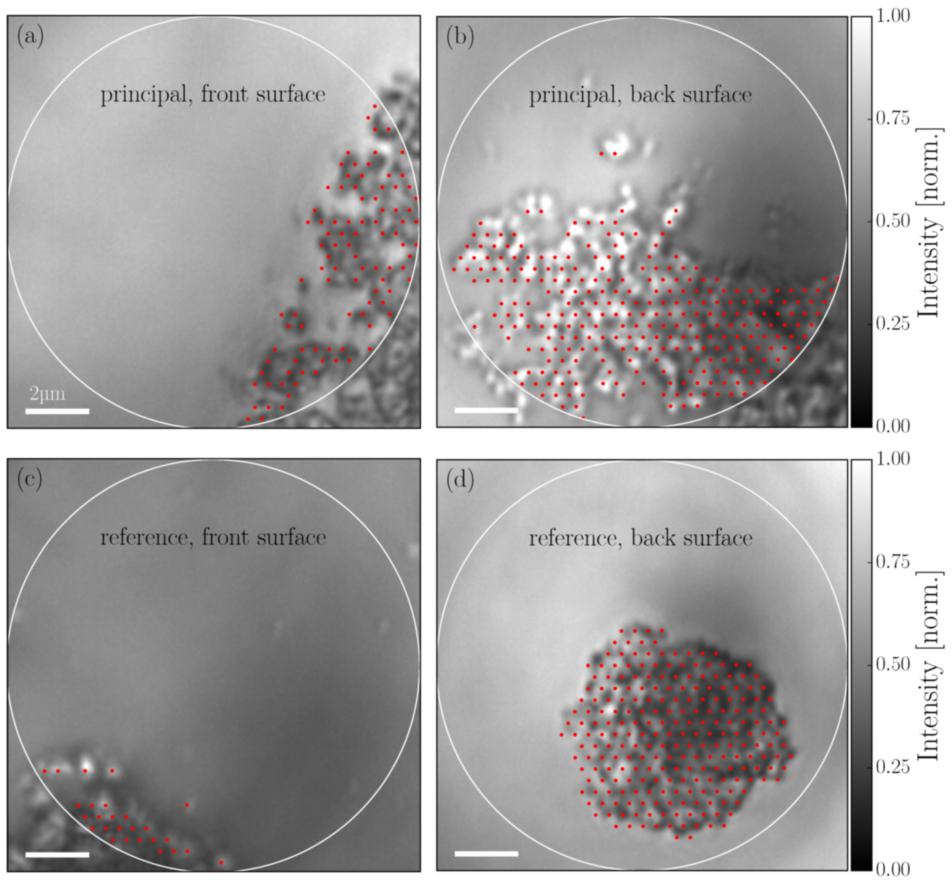}}
	\caption{Sparse sample used to study scattering avoidance. White light illumination of (a,c)~front and (b,d)~back surfaces of the sparse principal and reference media respectively. The white circle demarcates the region where the TM is measured. The red dots indicate the sampled grid points deemed to contain scatterers.
	}
    \label{fig:LED_images}
\end{figure}

\section{Scattering avoidance in sparse samples}

We now address the properties of SIMs inside   scattering media containing a sparse distribution of scattering particles.. In particular, we demonstrate that SIMs minimize the interaction of the light field with the scattering centers if  the sample is sparse enough to allow some light to pass unscattered. We  generate the SIMs based on the information stored in the transmission matrix. To be specific, we start with the conjecture that SIMs with a SIM-eigenvalue $\alpha \approx 1$ are exactly such scatterer-avoiding fields since they deliver the very same output field (in shape, amplitude and phase) when passing through the medium and through the reference.
To investigate whether there is indeed a corresponding correlation between $\alpha$ and the overlap of the wavefunction with the scatterer positions, we perform numerical simulations and experiments on samples with a sparse distribution of scatterers, where this overlap can be monitored. To demonstrate in parallel that our approach is applicable even in situations where a clear reference sample may be unavailable, we use a second sparse scattering sample as a reference.
In this case SIMs with $\alpha \approx 1$ are incident states that produce the same output patterns for both the principal and the reference scattering sample.

Our sparsely scattering samples are fabricated out of two layers of ZnO scattering particles separated by a transparent layer. Each scattering layer may be approximated as a single scattering system.
The thickness of the transparent layer is $\SI{15\pm1}{\micro\meter}$, which is much larger than the depth of field of our microscope objectives. Hence diffraction makes the field profiles at either face of the layer completely different. Importantly, we can image the two scattering layers separately under incoherent illumination as only one layer can be in focus at any time.
We use only a single polarization component since  polarization mixing is weak in systems without high order multiple scattering or surface roughness~\cite{banon2020depolarization}.

As a reference we use a different part of the same layer with an average coverage of 15\%. The corresponding TMs are resampled
in real space using a basis of Bessel modes of the first kind~\cite{Pai21},
which facilitates the determination of overlap of the fields with the scatterers in each layer.

The front and back surfaces of the principal sample have a scatterer coverage of 11\% and 32\% respectively, averaging to  22\%.

The SIM intensities at the two interfaces are computed from the measured TMs, allowing us to investigate the overlap of the light wave with the scatterers. The spatial intensity profile of the SIM that avoids the scatterers the best is depicted in the insets of Fig.~\ref{fig:GEV_avoidance}(a,b). We note that the field is concentrated in areas where no scatterers are present in either the principal or the reference sample. While being very similar in the two samples, the field is completely different on the front and back surfaces.

The main panels in Fig.~\ref{fig:GEV_avoidance}(a,b) display the statistics of SIMs and their corresponding intensity on the scatterers: the SIMs at high values of $|\alpha|$ diffract around the scatterers in both samples and hardly interact with them, i.e., they are almost purely ballistic light. At intermediate values of $|\alpha|$ the SIMs transmit by construction into identical patterns, even after interacting with the scatterers. At $|\alpha|$-values near zero the behavior becomes dominated by noise. 

The positions of the scattering parts of the samples were determined from white light illumination images which  are depicted in Fig.~\ref{fig:LED_images}.

Corresponding results obtained from numerical simulations with similar surface coverage and scatterer sizes are plotted in Fig.~\ref{fig:GEV_avoidance}(c,d). In the numerics we clearly observe a {very} similar trend where the states with the highest $|\alpha|$ avoid the scatterers the most, while more interaction takes place for states with lower $|\alpha|$. The interaction for the states with the highest $|\alpha|$ is much closer to zero than found in Fig.~\ref{fig:GEV_avoidance}(a,b). This can be attributed to the limited resolution of the images that are used to evaluate the overlap with scatterers in the experiment, causing us to overestimate the overlap.
The shape of the experimental and numerical distributions are sensitive to the fractional area coverage, size and clustering of the scatterers.

\medskip
\noindent \textbf{Double layer sample preparation}. The double layer scattering sample used in our experiments consists of two sparse layers of ZnO nanopowder separated by a thin film of optical glue. To prepare such a sample, we use the following recipe. We first airbrush a sparse semitransparent layer of ZnO particles (\SI{200}{\nano\meter} average size) on a plasma-cleaned microscope cover glass and dry it under ambient conditions.
Next, we spin coat a layer of optical glue with a thickness of around \SI{10}{\micro\meter}, after which we cure it with a UV-gun for \SI{3}{\min}. 
Finally, we spray a second layer of ZnO on top of the cured glue and dry it.

For the numerical simulation of the double-layer structures in Fig.~\ref{fig:GEV_avoidance}, we model the optical glue film with a polygonal scatterer with refractive index $n_\text{glue} = 1.5$ whose length is again adjusted according to the experimental values, i.e., $L_\text{glue} = \lambda (L_\text{glue}^\text{exp}/\lambda^\text{exp})$ with $L_\text{glue}^\text{exp} \approx 15\ \mathrm{\mu m}$. Since this film shows some thickness variations in the experiment, we incorporate those by adding random variations $\delta L \approx r \lambda$ with $r$ being a random number between 0 and 1, to 25 equally spaced points along the width of the film on both sides of the polygon. Since the scatterers at the output side are immersed in the optical glue in the experiment, we also take this into account by placing the circular scatterers at the output side inside our polygon. Since the experimental scatterer distribution shows a clustering of the ZnO nanoparticles, we also create a clustered configuration by drawing random scatterer positions near the surface of our glue polygon around two cluster centers which we also randomly choose. Each scatterer size again occupies the same fraction of the surface coverage of the corresponding layer and sample. 

To calculate how much intensity of a SIM falls onto the scatterers, we use a method similar to the one used in the experiment, where we transversally read out the SIM wavefunctions right in front of (after) the  scatterers on the input (output) layer and integrate the corresponding intensities at the position of the scatterers. 
\begin{figure*}
\centering
\includegraphics[width=0.8\textwidth]{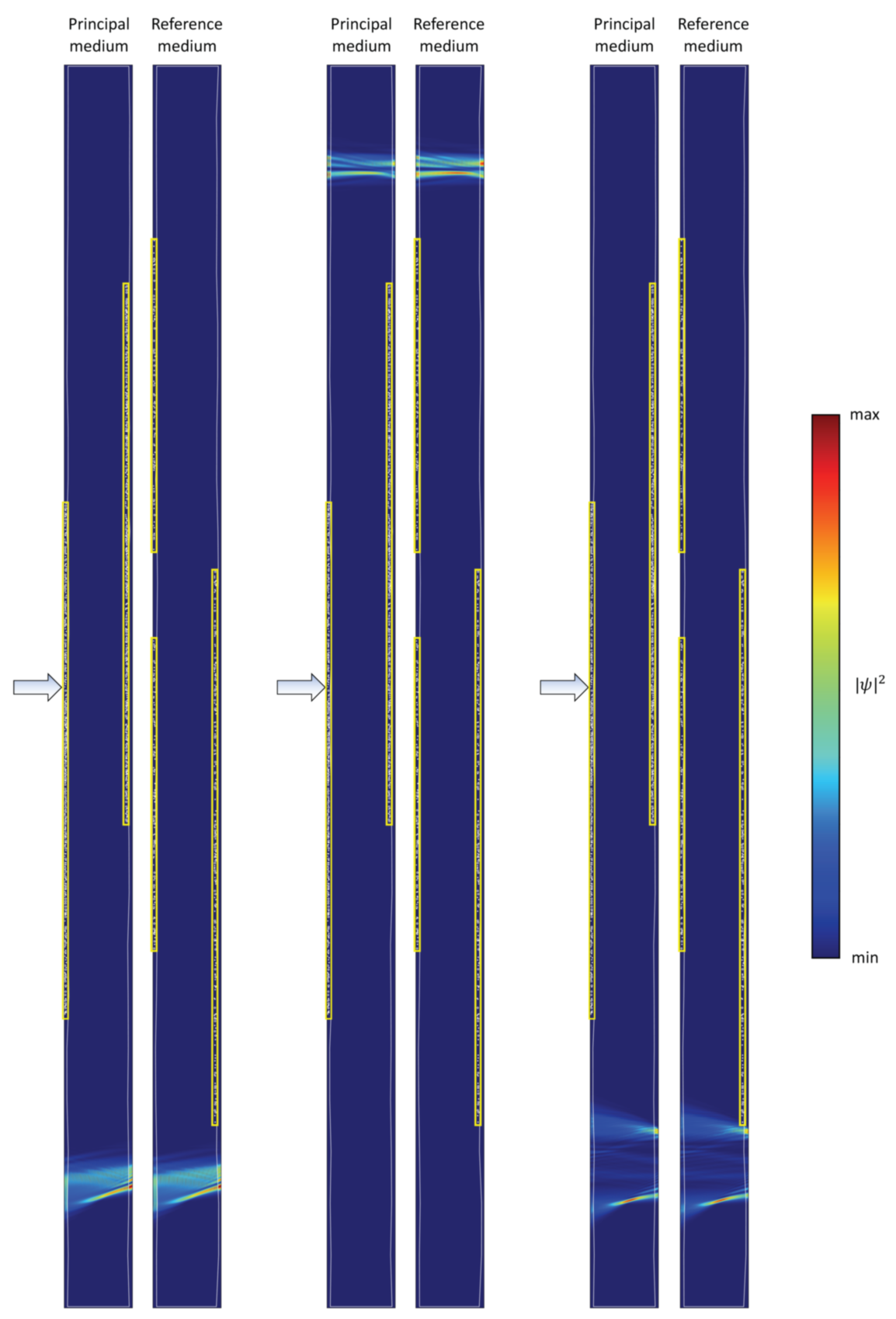}
\caption{Simulated intensity distributions of SIMs with a high $|\alpha|$, where each of these states is propagated through the principal and the reference medium. The arrow marks the input port, whereas the yellow rectangles highlight the position of the clustered scatterers.}
\label{fig:SIM_numerical_wavefunctions}
\end{figure*}
\medskip

\noindent \textbf{Simulated SIM wavefunctions}. In Fig.~\ref{fig:SIM_numerical_wavefunctions} we show three SIMs of the simulated double layer structure in Fig.~\ref{fig:GEV_avoidance} which feature a high $|\alpha|$ (see Methods for details of the simulations). The latter causes the output profiles to be almost identical, whereas the property of scatterer avoidance in such sparse samples yields a high degree of similarity of the entire intensity distributions inside the principal and the reference medium. Note that even though these states spare out the scatterers, they still propagate through the polygonal scatterer which models the optical glue. Since this scatterer features thickness variations which are different for the two samples, the intensity distributions are not identical, but very similar.

\section{Numerical simulations of imaging internal objects in 3D}
\label{sec:Imaging_3D}

In this section we provide detailed information on how we carry out the 3D wave simulation and illustrate the potential of SIMs for imaging in complex media.

\subsection{Generation of transmission matrices}

\begin{figure}[h!]
\centering
\includegraphics[width=0.85 \columnwidth]{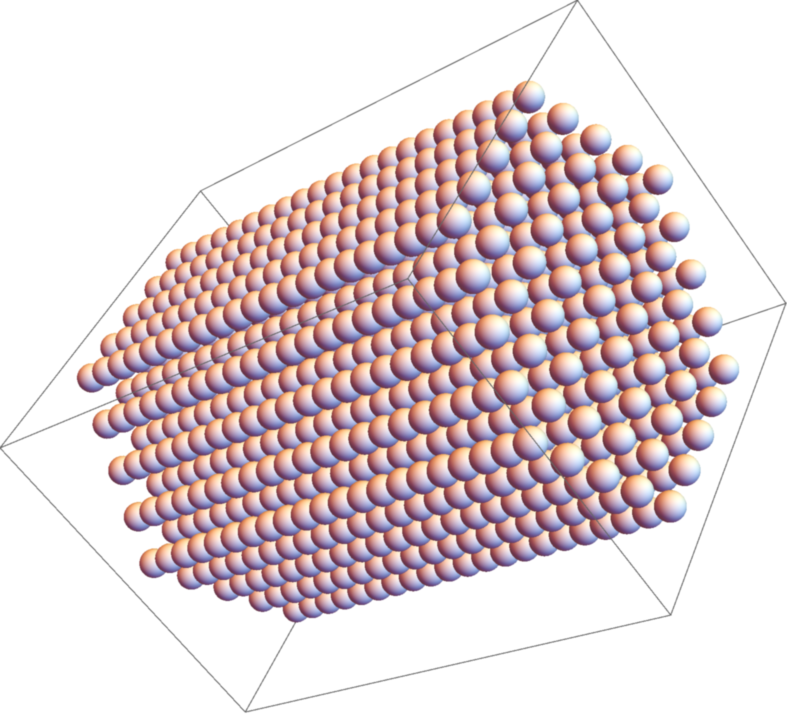}
\caption{Sketch of the hexagonal lattice used for the 3D wave simulations.}
\label{fig:hexagonal}
\end{figure}

To efficiently generate an ensemble of 3D transmission matrices we computationally solve an Anderson-like lattice model using a recursive S-matrix method. The non-scattering lattice supports the propagation of scalar waves with a dispersion relation that is close to that of free space. The lattice is a 3D hexagonal lattice, representing a waveguide with a hexagonal cross section (see Fig.~\ref{fig:hexagonal}), consisting of stacked hexagonal planes with 1801 lattice points per plane.

The use of a hexagonal lattice makes the diffraction of a beam in the empty waveguide less anisotropic than would be the case for a square lattice. 
The advantage of a lattice hopping model over a true finite-difference model such as employed in finite-difference time-domain (FDTD) simulations is that the model is stable even for fairly large values of the lattice constant up to the order of half a wavelength.

For an empty section of the waveguide the transmission matrix is found by diagonalizing the 2D Laplacian matrix
\cite{kantorovich} and by propagating the solutions in the longitudinal $z$-direction according to the spectral method
 $\Psi(z)=\sum_i \Phi_i  e^{i \beta_i z} \langle \Phi_i| \Psi(0)\rangle$,
 where the $\Phi_i$ are the eigenvectors of the 2D Laplacian and $\beta_i$ are the propagation constants, with $\beta_i^2+\omega^2=\Lambda_i$, where $\Lambda_i$ is the corresponding eigenvalue of the 2D Laplacian matrix and $\omega$ is the angular frequency.
We model scattering by considering a zero-thickness section of the waveguide that contains point scatterers at certain lattice points. The transmission matrix of such a zero-length section is diagonal in a lattice point basis. Its diagonal elements corresponding to positions where no scatterer is present are $t_{ii}=1$, and the corresponding reflection matrix element is zero. The transmission matrix elements corresponding to the position of a scatterer are given as $t_{ii}=1-v+i \sqrt{v-v^2}$ and the corresponding reflection matrix element as $r_{ii}=-v+i \sqrt{v-v^2}$, where $-1<v<1$ is a parameter controlling the scattering strength. Note that this formulation captures the phase shift in scattering as well as the backscattering property of a dielectric scatterer, while ensuring that the $S$-matrix of the scattering system remains unitary.

The scattering matrix of a complete sample is then built up by iterative use of the S-matrix composition rule~\cite{Ko1988}, stacking alternately short free-propagation sections and zero-length scattering sections, while taking into account multiple internal reflections in an exact way. The addition of a propagation step requires only a few matrix multiplications, whereas the addition of a scattering step entails inversion of a $N_{2d} \times N_{2d}$ matrix to take into account the multiple reflections. 

Once the sample is built up in this way (based on a sample generation algorithm coded in Mathematica), we attach short empty waveguide sections (``leads'') on both sides to prevent artefacts due to the nearby presence of a scatterer.  For each parameter set of the waveguide (scattering strength, scatterer density and length), tens of matrices are generated to obtain sufficient statistical significance. 

The  field at a depth $0<z<L$ inside the scattering sample is $E(x,y,z)$, which can be found in a vector representation from the transmission and reflection matrices of the left and right layers as
\[
E_{\rm scat}(z)=  (1+R^{L}_B+R^{R}_A R^{L}_B+R^{L}_B R^{R}_A R^{L}_B +...)T_A E_{\rm in},
\]
where $T_A$ is the transmission matrix of the left layer ($A$) of the sample, $R^{L}_B$ is the left-to-left reflection matrix of the right layer ($B$) of the sample, and $R^{R}_A$ is the right-to-right reflection matrix of the left part of the sample. The infinite (geometric) series can be summed as follows
\[
E_{\rm scat}(z)=  (1+R^{L}_B)(1-R^{R}_A R^{L}_B)^{-1}T_A E_{\rm in}.
\]
Thus, we calculate the field at depth $z$ including directly transmitted as well as multiply reflected terms, which in the center of strongly scattering samples may dominate the direct term.

Finally, we obtain the SIMs from the overall system transmission matrix $T_s$  by decomposing $T^\dag_{\rm air}(L) T_s=A \alpha A^{-1}$, where $\alpha$ is a diagonal matrix, $T_{\rm air}(L)$ is the transmission matrix of the homogeneous reference medium (air) of the same thickness $L$ as the scattering sample, and $A$ is the (in general non-unitary) matrix of SIMs. 
To correct for the overall phase shift caused by the effective refractive index of the scatterers, we rotate the overall phase by $e^{- i \phi_0}$, where $\phi_0$ is the average phase of the ballistic waves, $\phi_0={\rm Arg}({\rm Trace}\, T_s)$. This overall phase rotation ensures that the SIMs with phase zero are in phase with the ballistic waves.

In the case of a very thick or very strongly scattering sample (more than about 10 scattering mean free paths $\ell$) or of a large mismatch between the index of the reference medium and the effective index of the scattering medium, the overall phase rotation cannot be determined straightforwardly. We therefore consider here only the case of samples of thickness $L<10 \ell$, and assume that the reference medium is closely (but not necessarily perfectly) index-matched to the effective index of the scattering medium.

\subsection{Fidelity of SIMs inside the medium}
The defining property of SIMs is that they exit the sample with the same field profile as the ballistic light, albeit multiplied by the complex SIM eigenvalue $\alpha$. We hypothesize that their field profile $E(z)$ at a depth $0<z<L$ inside a sample of thickness $L$ remains correlated with the ballistic field profile as long as the sample is not too thick. 

To illustrate this point, let us consider first the SIM eigenvalues of a very weakly scattering sample (much thinner than a mean free path), which all lie close to the ballistic value of $1+0 i$. As the scattering strength is gradually increased, the eigenvalues diffuse away from that point and distribute themselves in a gradually more symmetric distribution around the complex origin. In 2D simulations we are able to track the eigenvalues as a function of scattering strength (see Fig.~\ref{fig:SIM_eval_evolution} and the associated supplemental video file). 

For SIMs with eigenvalue $\alpha \approx 1$ in an optically thin medium, the propagation inside is essentially ballistic, as they interact very little or not at all with the scattering centers, and hence the field profile inside the scattering medium correlates  well with that in the reference medium, as seen in Fig.~\ref{fig:SIM_numerical_wavefunctions}.
We hypothesize that for SIMs in media of much larger optical thickness, their field profile in the {\em scattering}\ medium remains similar to that of the same wave propagated in the {\em reference}\ medium. We check this hypothesis numerically by calculating the  fidelity $F$ for a large number of SIMs according to 
\begin{equation}
F= \frac{|\langle E_{\rm ref}(z)|E_{\rm scat}(z) \rangle|^2}{\langle E_{\rm scat}(z) |E_{\rm scat}(z) \rangle\langle E_{\rm ref}(z) |E_{\rm ref}(z) \rangle}. 
\label{eq:ampF}
\end{equation} 
Here, $E_{\rm ref}(z)$ represents the field of the normalized SIM propagated through the reference medium and $E_{\rm scat}(z)$ is the field inside the scattering medium of the same SIM. The brackets denote an inner product by integration over the $x,y$ coordinates.

\begin{figure}[t!]
\centering
\includegraphics[width=0.85\columnwidth]{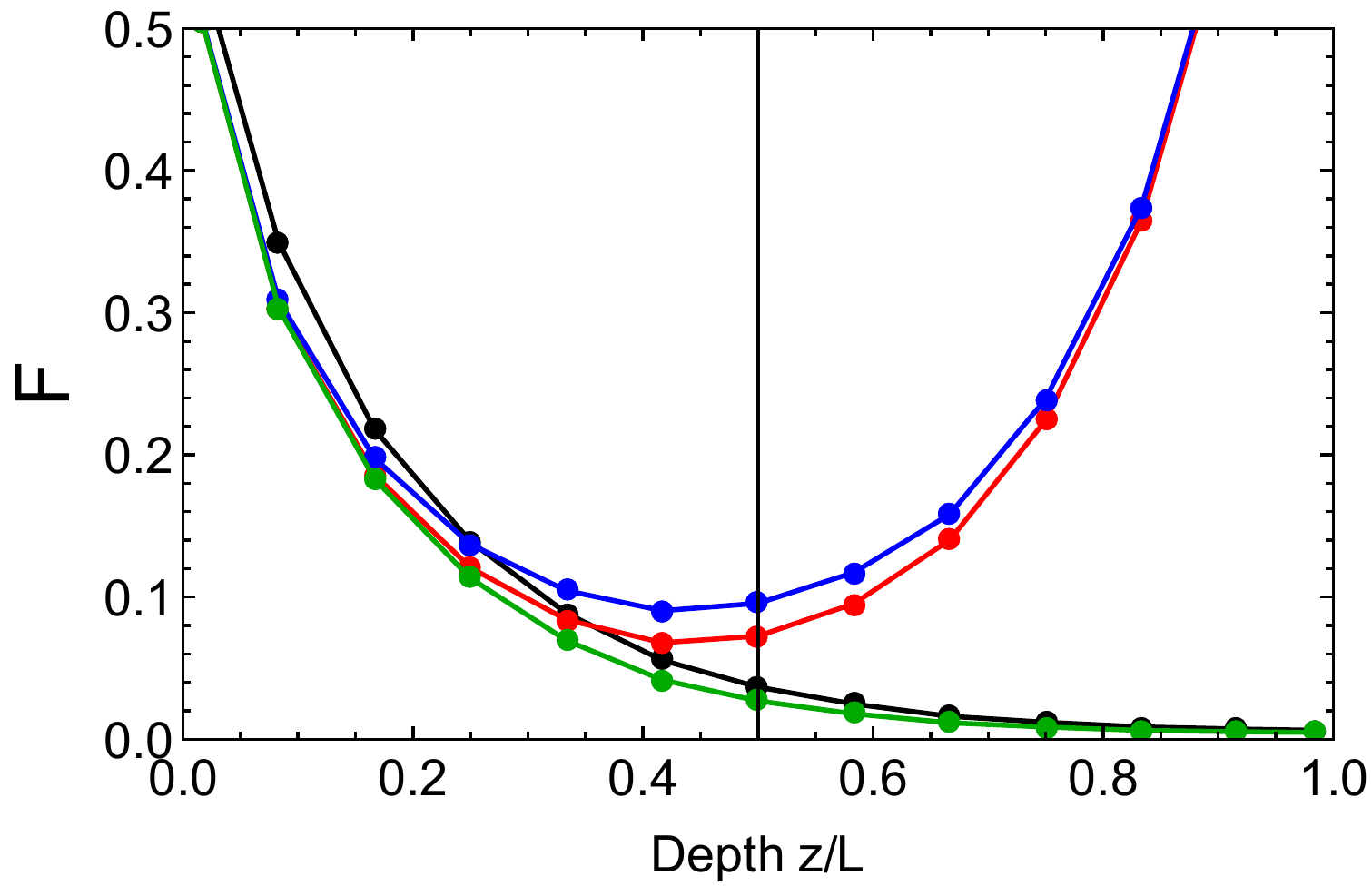}
\caption{Fidelity $F$ versus depth $z$ inside a scattering medium of thickness $L=7\ell$, as calculated for waveguide modes (black dots), an average over all SIMs (red) and SIMs of phase close to 0 (blue). Also shown is the fidelity for eigenstates of the transmission matrix (green).}
\label{fig:fidelityvsdepth}
\end{figure}
In Fig.~\ref{fig:fidelityvsdepth} we show the average fidelity versus depth in a medium of thickness $L=7 \ell$, as calculated for incident waveguide modes (black dots), an average over all SIMs (red dots), and an average over the subset of SIMs with $-\pi/2<\phi<\pi/2$ (blue dots). 
As expected, the fidelity of the reference waves (waveguide modes) decays approximately exponentially with depth. However, the fidelity of SIMs has a markedly different behavior, decaying to a minimum approximately in the center of the medium and thereafter rising again. The increase of the  fidelity of SIM waves at $z>L/2$ can be understood from the fact that at $z=L$ the field profile of a SIM is by construction very close to the field propagated through the reference medium, hence after propagating through the entire sample they should correlate very strongly with the ballistically propagated wave.

One may wonder whether this enhanced fidelity is special to SIMs or a rather general feature of any reasonable decomposition of the transmission matrix such as its regular eigenvector decomposition. 
To test this we also show in Fig.~\ref{fig:fidelityvsdepth} the fidelity as calculated for eigenvectors of the TM (green dots).  We observe that deep inside the sample, eigenvectors show a fidelity that is very similar to that of random modes and much lower than that of SIMs. We note that on the center of the sample the fidelity of eigenvectors corresponding to high eigenvalues is about 20\% higher than that of low eigenvectors (these curves are not separately plotted because the difference is too small to be clear in the plot). We observed identical behavior for singular vectors of the transmission matrix. This leads to the conclusion that the high fidelity is a specific property of SIMs that is not shared by the other decompositions of the TM.

We observe that for positions $z \approx L/2$ the SIMs inside the medium are characterized by a higher fidelity than the incident waveguide modes.
In addition, the SIMs with $-\pi/2<\phi<\pi/2$ clearly have a higher fidelity than the average SIMs, confirming the relation between the SIM phase and fidelity.
Importantly, in the center of the medium, the fidelity of the averaged SIMs is about twice that of the reference waves, while for the SIMs with phase $-\pi/2<\phi<\pi/2$ the fidelity is about 3 times higher than the reference.

Knowledge of the correlated transmission of the complex, speckle-like SIM waves can be useful for methods such as compressive imaging. When one is interested in superpositions of SIMs to construct, e.g., a focus inside the scattering medium, in addition knowledge of the overall phase the wave accrues is essential. 
For a SIM with phase $\phi=\arg(\alpha)=0$ it is reasonable to expect that the phase of the wave inside the scattering medium is the same as that of the wave in the reference medium. In case $\phi \ne 0$
one may expect that this additional phase is accrued linearly with $z/L$.
\begin{figure}[tb]
\centering
\includegraphics[width=0.85\columnwidth]{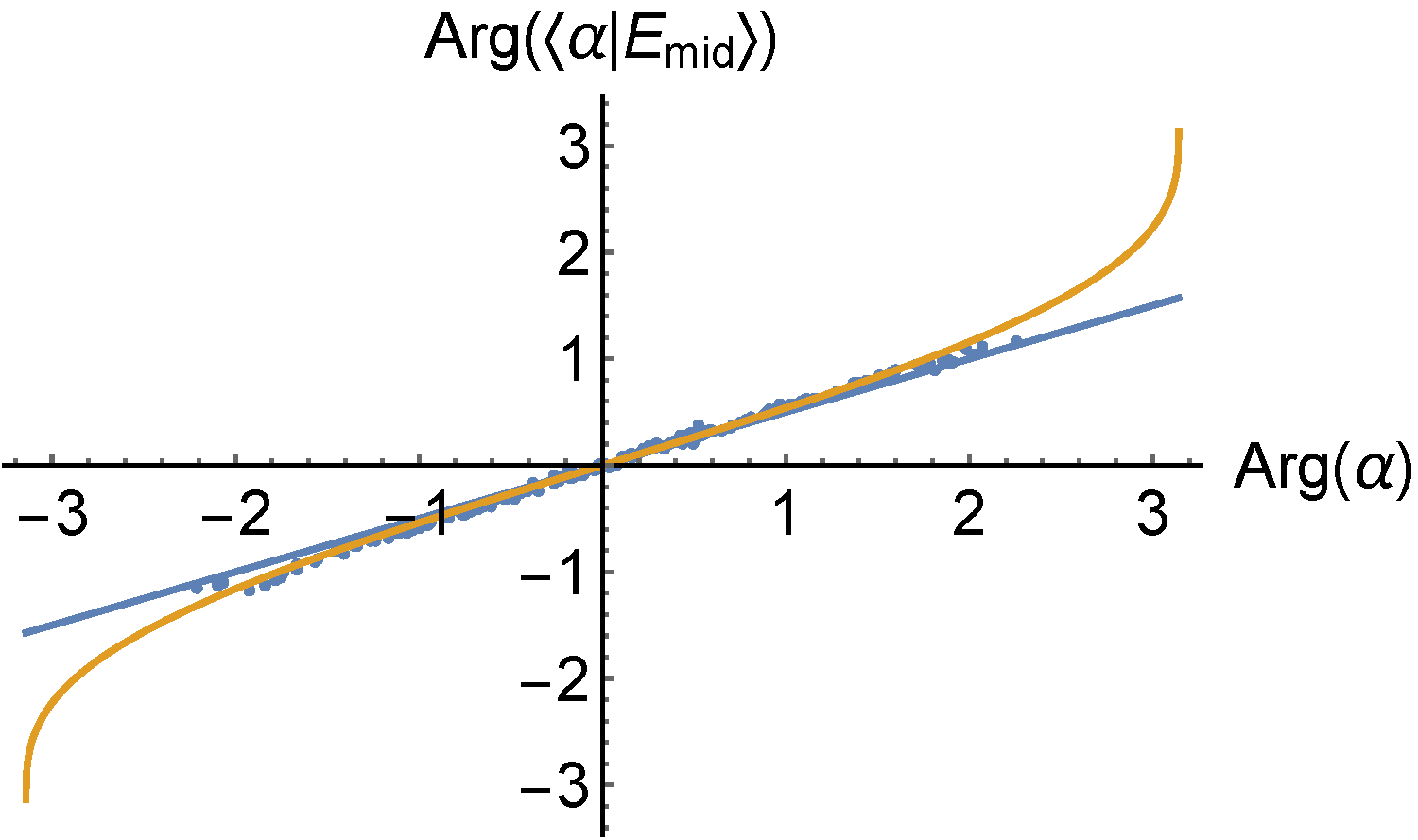}
\caption{Phase of the SIM in the object plane vs phase of the SIM eigenvalue, for a weakly scattering system (system thickness $L=2.6\, \ell$). Blue line: $y=x/2$. Yellow curve: heuristic approximation, see equation~\ref{eq:heuristicphase}.}
\label{fig:argargweakscattering}
\end{figure}
\begin{figure}[h!]
\centering
\includegraphics[width=0.85\columnwidth]{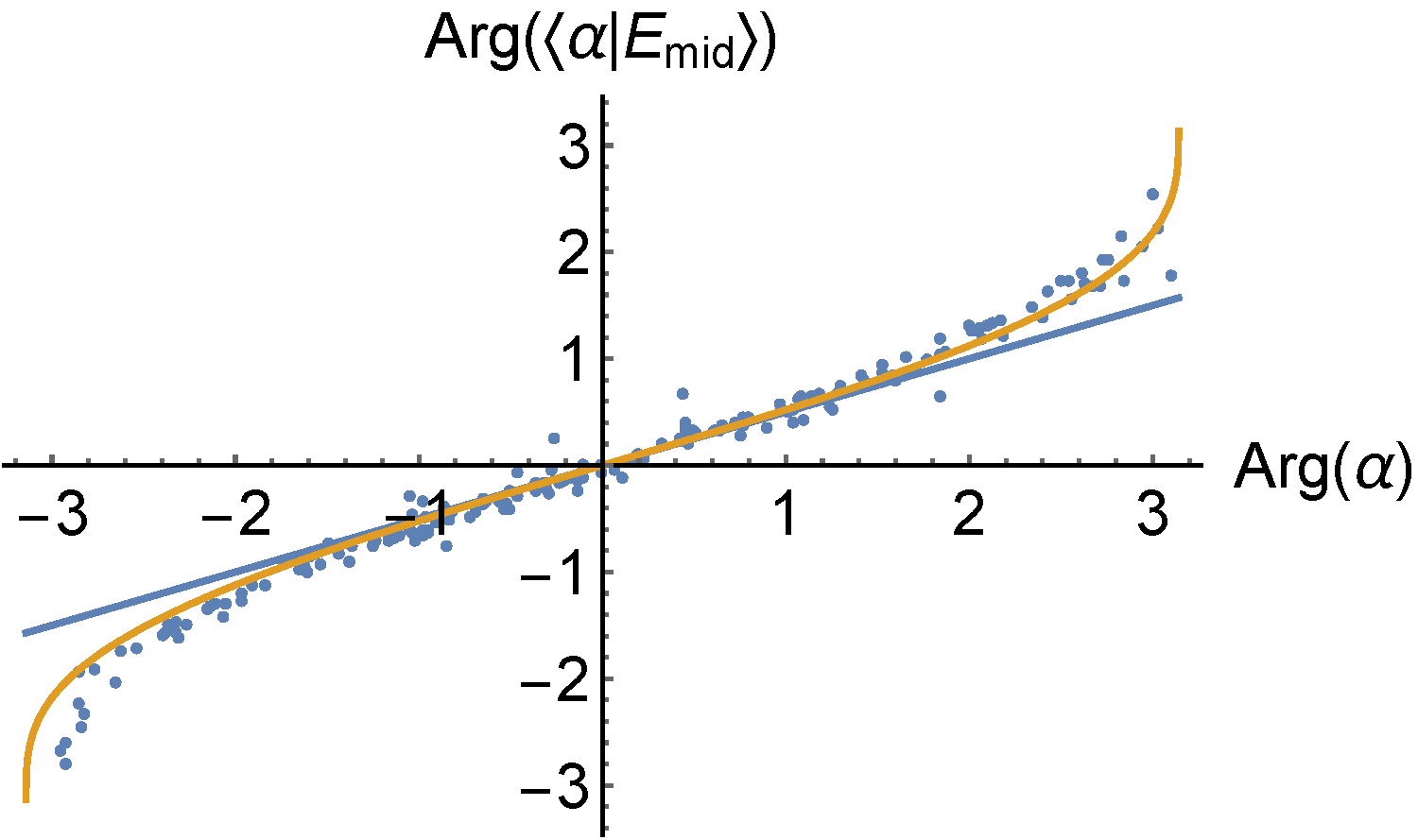}
\caption{Phase of the SIM in the object plane vs phase of the SIM eigenvalue, for a more strongly scattering system (system thickness $L= 7.8\, \ell$). Blue line: $y=x/2$. Yellow curve: heuristic approximation, see equation~\ref{eq:heuristicphase}.}
\label{fig:argargf40}
\end{figure}

In Fig.~\ref{fig:argargweakscattering}
we show the phase of SIMs at depth $z=L/2$, denoted as
$\arg(\langle E_\alpha(z) | E_{\rm scat}(z) \rangle)$,
where
\[
\langle E_\alpha(z) | E_{\rm scat}(z) \rangle \equiv 
(T_{\rm air}(z) E_\alpha)^\dag \cdot E_{\rm scat}(z)\,.
\]
Here $T_{\rm air}(z)$ represents the transmission matrix of a  reference medium with thickness $z$, and $E_\alpha$ is the  field of a SIM on the input side of the medium.
It is seen that the phases of SIMs at $z=L/2$ lie close to the line $y=x/2$, which fits with the naive expectation that, indeed, about half the phase shift is accumulated in the first half of the sample.
For a more strongly scattering system the same plot is shown in Fig.~\ref{fig:argargf40}. Here  the  eigenvalues cover  the full range of phases $(-\pi,\pi)$. There is a clear deviation from the linear behavior at the ends of the interval, which is well described by the heuristic curve
\begin{equation}
    \psi=\pi\frac{(\phi +\pi)^{m}-(\pi-\phi)^{m}}{(2 \pi)^{m}},
    \label{eq:heuristicphase}
\end{equation}
where $\psi=\arg(\langle\alpha | E_{\rm mid} \rangle)$ is the phase of the field inside, $\phi=\arg(\alpha)$, and $m$ is a parameter that depends on the depth of the imaging plane inside the medium, where we found reasonable heuristic agreement for $m=(z/L)^{3/2}$.

In conclusion, the field of a SIM inside a reference medium resembles the field inside the scattering medium with a fidelity that can be considerably higher than that of plane waves or waveguide modes. The relative phase between different SIMs is accurately predicted by a simple heuristic formula, which allows one to use superpositions of SIMs to construct arbitrary fields such as foci inside a random medium.

\subsection{3D Deep imaging simulation}
The ability to construct superposition fields inside the random medium suggests that SIMs can be used to improve imaging of objects deep inside the medium.
We simulate an experiment where a 2D, very weakly fluorescent object is present in the most challenging position inside a 3D scattering medium, namely in the center at $z=L/2$. We choose the shape of the object to resemble a constellation as shown in Fig.~4b of the main text.

 The medium in which this object is embedded in our simulation is strongly scattering, comparable to the ZnO media used in the experiments, with $\ell \approx 2 \lambda$. The medium thickness is  $L \approx 7 \ell$. 
As a reference method, we simulate a scanning fluorescence microscopy experiment, where a single-frequency excitation beam is focused in the plane of the object and translated while the total fluorescence is recorded.

In case of uncorrected scanning fluorescence microscopy, a beam is incident on the sample of which the ballistic component is focused to a field $E_{\rm foc}(x,y)$ at depth $z$. The corresponding incident field is given by $E_{\rm in,bal}=T_{\rm air}^\dag(z) E_{\rm foc}(x,y)$, where $T_{\rm air}(z)$ is the transmission matrix of the homogeneous reference medium (air) of thickness $z$.

Using the decomposition of the transmission matrix into SIMs we aim to construct an incident field that provides a better focus inside the scattering sample than $E_{\rm in, bal}$. To do so we first consider the fidelity of the SIMs at the location $z$, which is assumed to be near the center of the sample, $z \approx L/2$.

\begin{figure}[t!]
\centering
\includegraphics[width=0.85\columnwidth]{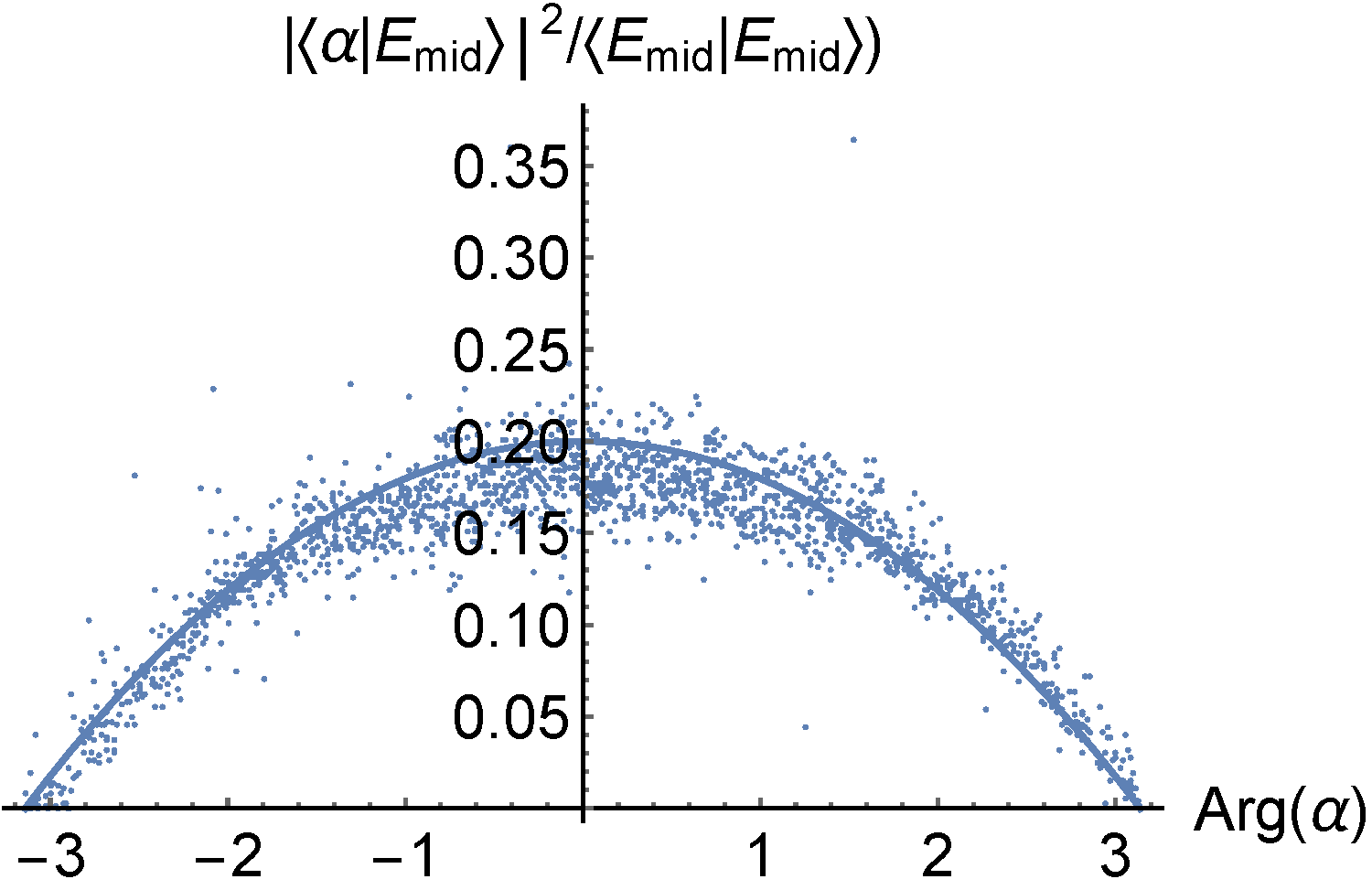}
\caption{Fidelity of the SIMs in the center of the sample plotted versus the phase of their eigenvalue, for a sample with system thickness $L=7.1\, \ell$. The parabolic curve is a heuristic description, see equation~\ref{eq:fidelity}. }
\label{fig:FidelityArg}
\end{figure}

We plot the fidelity of each SIM versus the phase of the SIM eigenvalue in Fig.~\ref{fig:FidelityArg}.
Remarkably, the fidelity strongly depends on the phase $\phi$ of the SIM, heuristically described by a parabola, 
\begin{equation}
    F \propto \left[1-(\arg \alpha/\pi)^2\right].
    \label{eq:fidelity}
\end{equation} 
Since the fidelity of the SIMs with phase close to $\pm \pi$ is essentially zero, it is preferable to use SIMs with phase close to zero to reconstruct a focused field. Heuristically, weighing the contribution of each SIM by its anticipated fidelity according to equation~\ref{eq:fidelity} provides the best results.

Hence in order to produce a focused field $E_{\rm foc}(x,y,z)$ inside a scattering sample of thickness $L$, using only the experimentally accessible knowledge of the full system transmission matrix $T_s$, and the easily obtained transmission matrix $T_{\rm air}(z)$ of an air sample of thickness $z$, we apply the following procedure:
\begin{enumerate}
\item We obtain the SIMs by decomposing $T^\dag_{\rm air}(L) T_s=A \alpha A^{-1}$, where $\alpha$ is a diagonal matrix and $A$ is the (in general non-unitary) matrix of SIMs.
    \item We obtain the uncorrected incident field $E_{\rm in,bal}=T_{\rm air}^\dag(z) E_{\rm foc}(x,y)$.
    \item Then we decompose the back-propagated field into SIMs using the matrix $A^{-1}$.
    \item We estimate the phase correction $\psi$ and the fidelity $F$ from Eqs.~\ref{eq:heuristicphase} and \ref{eq:fidelity}, respectively.
    \item We multiply each SIM contribution by $e^{- i \psi}$ to correct the phase inside the sample.
    \item We multiply the amplitude by the estimate of $F$ (equation~\ref{eq:fidelity}).
    \item Finally, we compose the corrected incident field by multiplying the vector of corrected complex SIM amplitudes  with $A$, to obtain a corrected incident field $E_{\rm in,SIM}$.
\end{enumerate}
This procedure corrects the phase of each SIM component in the target field by assuming that they accrue approximately a fraction $z/L$ of the total phase while propagating through a thickness $z$ of the scattering medium (see Fig.~\ref{fig:argargweakscattering} and \ref{fig:argargf40}). In this way we place emphasis on those SIMs that are estimated to have a high fidelity in the object plane, in order to maximize the power in the focus.

\subsection{Results}

\begin{figure}[tb]
\centering
\includegraphics[width=0.85\columnwidth]{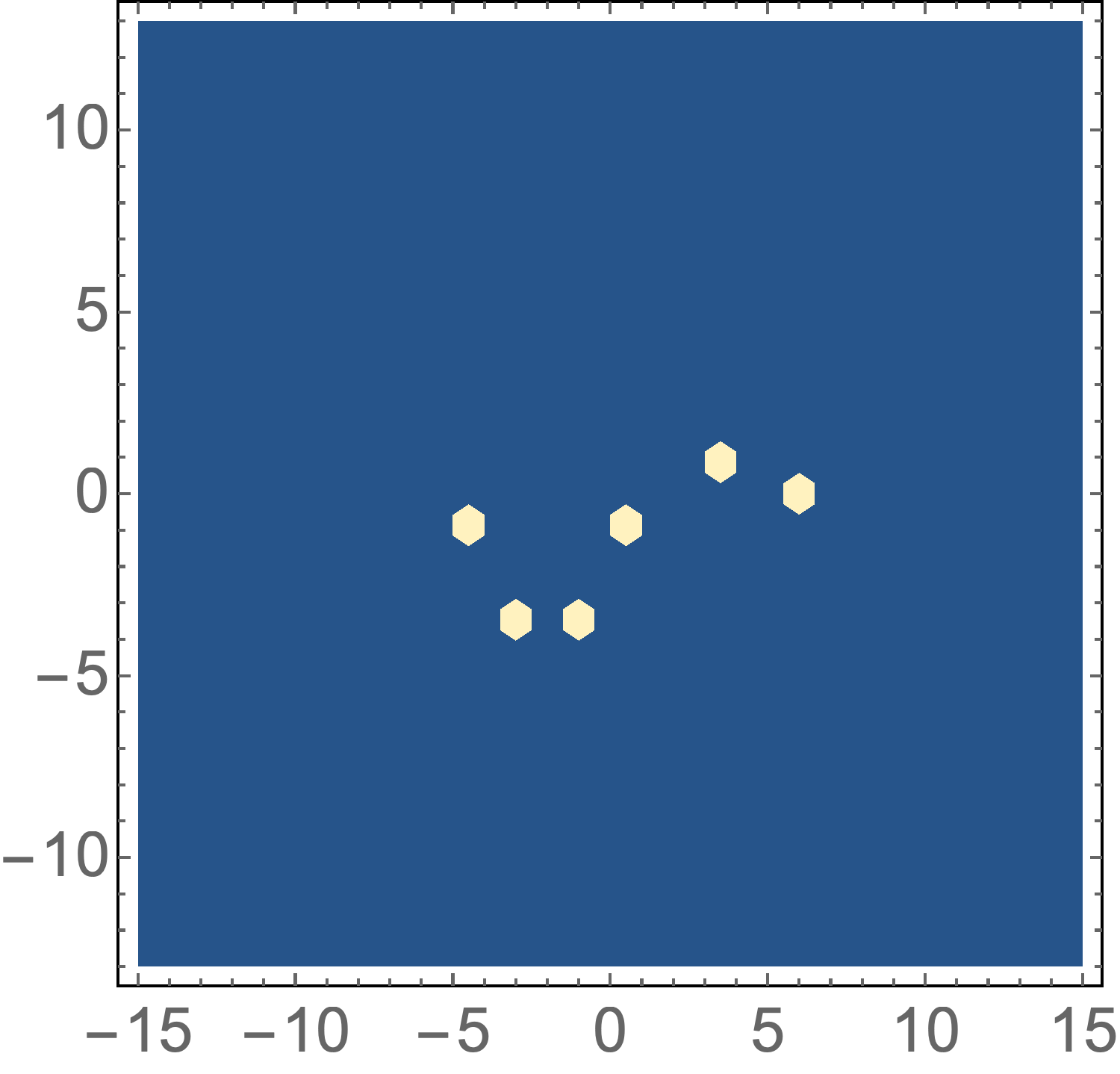}
\caption{The test object used for evaluating the imaging performance consists of 6 bright points on a dark background resembling a stellar constellation. }
\label{fig:object}
\end{figure}

We have simulated both standard fluorescence imaging (using $E_{\rm in, bal}$) and SIM-corrected imaging (using $E_{\rm in, SIM}$) for a range of parameters of the scattering layers. For simplicity we have always chosen the fluorescent object to be sandwiched between scattering sections from the same ensemble. The fluorescent test object that we use to evaluate the imaging performance is shown in Fig.~\ref{fig:object}.

The lattice and object dimensions are given in calculational units, where the lattice constant of the 2D hexagonal planes in the 3D waveguide is 1, the distance between the different planes is 0.5, and the wavelength is $\lambda= 2$. We note that in this 3D tight-binding model the wavelength is not much larger than the lattice constant, leading to a slight change in the dispersion relation. In the 2D simulations presented in the next section we obtain similar conclusions with a more realistic dispersion relation, indicating that our results do not depend critically on the dispersion relation.
Each hexagonal plane consists of 24 hexagonal rings around the central point and hence has 1801 lattice points (corresponding to  a diameter of $25\lambda$). 
The scatterer strength $v$ is chosen to be 0.25 and the scatterer density $n_s$, which is the fraction of lattice points in the scattering section that contain a scatterer, is varied to change the mean free path from 4 to 6 wavelengths.

\begin{figure}[tb]
\centering
\includegraphics[width=0.75\columnwidth]{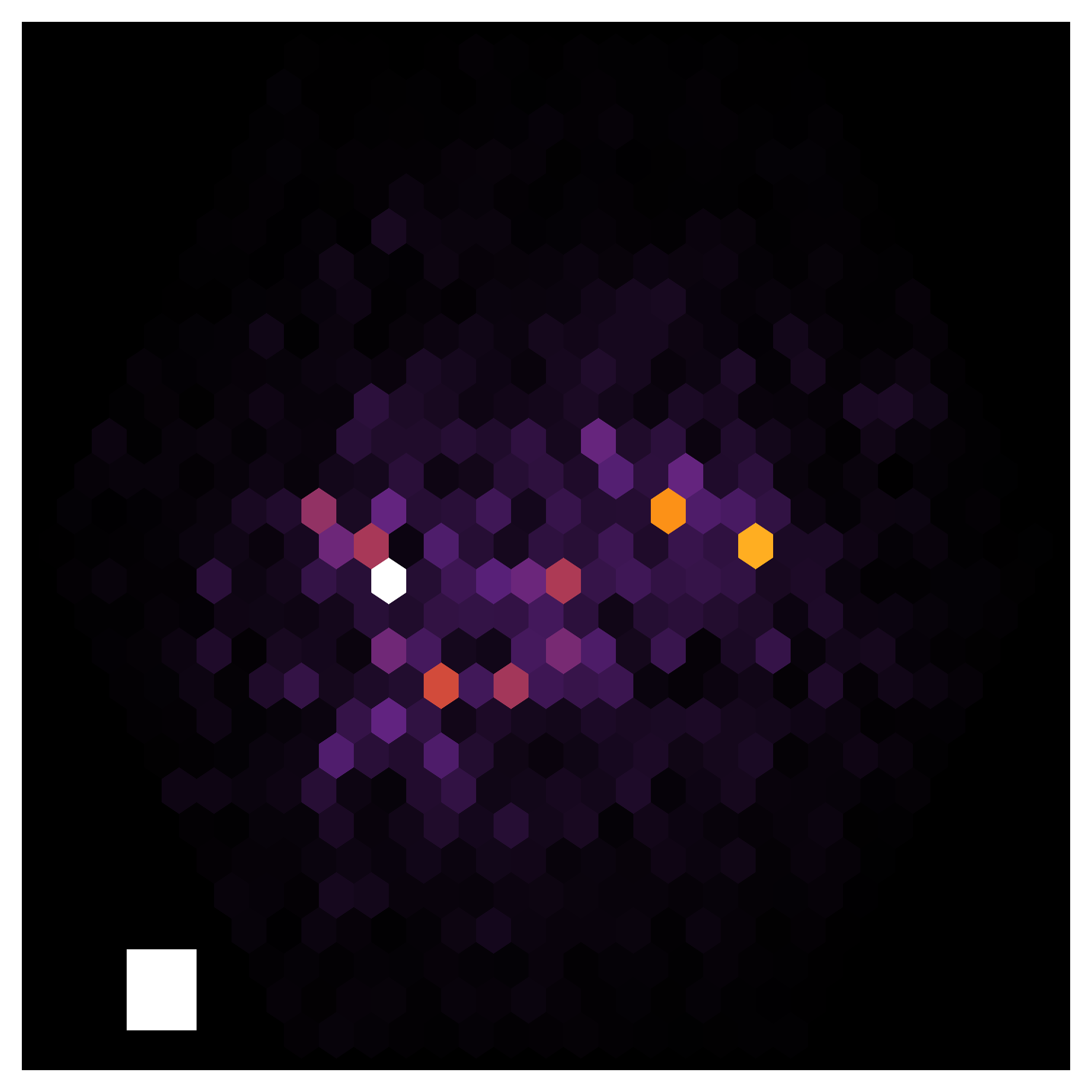}
\includegraphics[width=0.75\columnwidth]{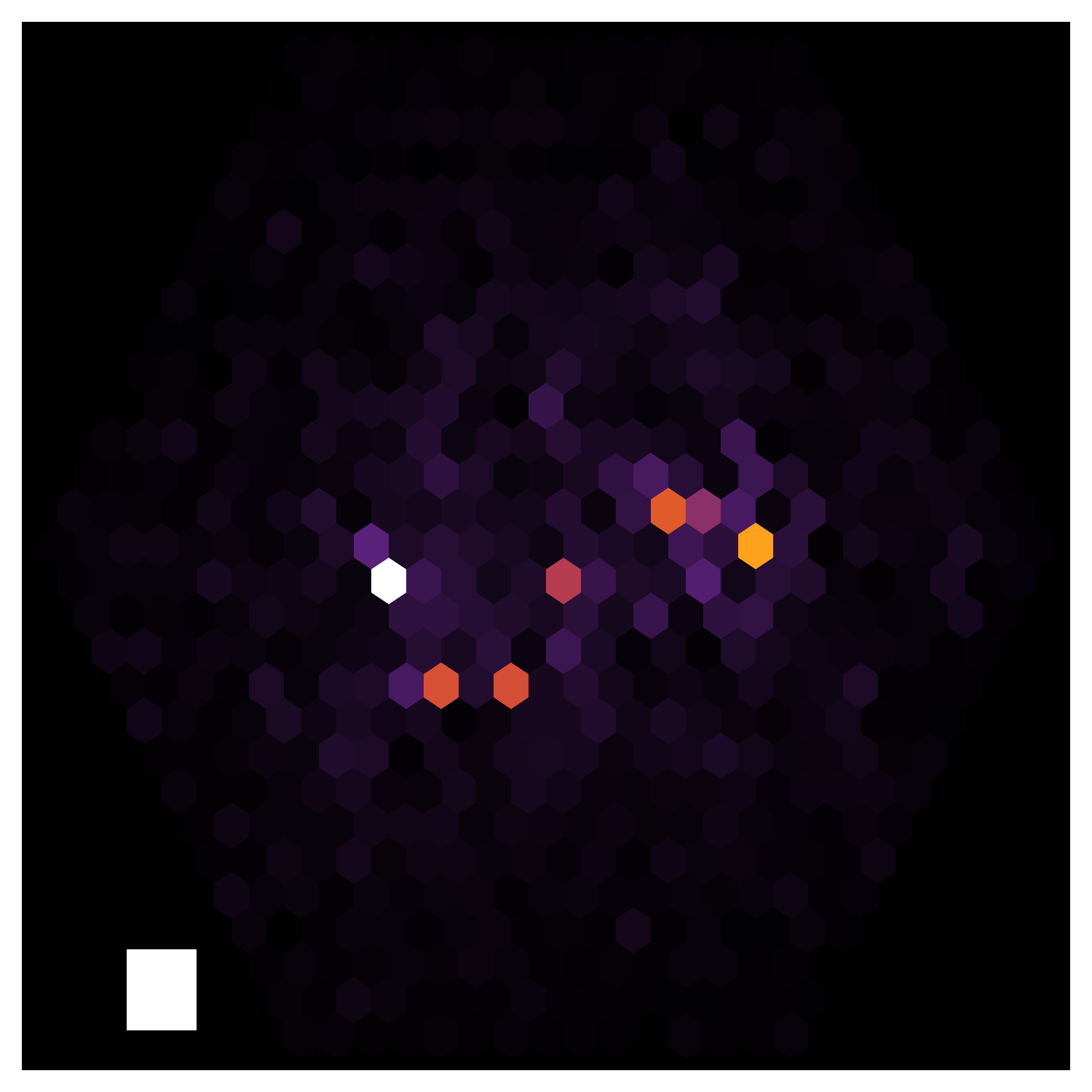}
\includegraphics[width=0.75\columnwidth]{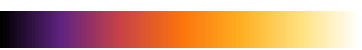}
\caption{Typical results at system thickness $5.4\, \ell$. Top: Uncorrected image (Pearson correlation value $C_{\rm uncorr.}=0.747$), Bottom: Corrected image (amplitude and phase correction, $C_{\rm SIM}=0.828$). The white squares indicate the scale of the wavelength in the simulations.}
\label{fig:typicalg20}
\end{figure}

For a sample of intermediate scattering strength (total optical thickness of $L/\ell=$ 5.4), we have plotted the simulated scanning excitation images before and after correction in Fig.~\ref{fig:typicalg20}. The object (as shown in Fig.~\ref{fig:object}) is already discernible in the uncorrected image, the corrected image shows a much lower noise level. This is also obvious from the (Pearson's) correlation value $C$ between the images and the true object, which in this example is 0.747 for the uncorrected case and 0.828 after correction. The correlation coefficients are calculated over the visible image in the figure. Including the far-out-of-focus regions leads to lower correlations both for the corrected and uncorrected images. In any realistic setting photons from these regions are excluded by a detection aperture.

\begin{figure}[h!]
\centering
\includegraphics[width=0.75\columnwidth]{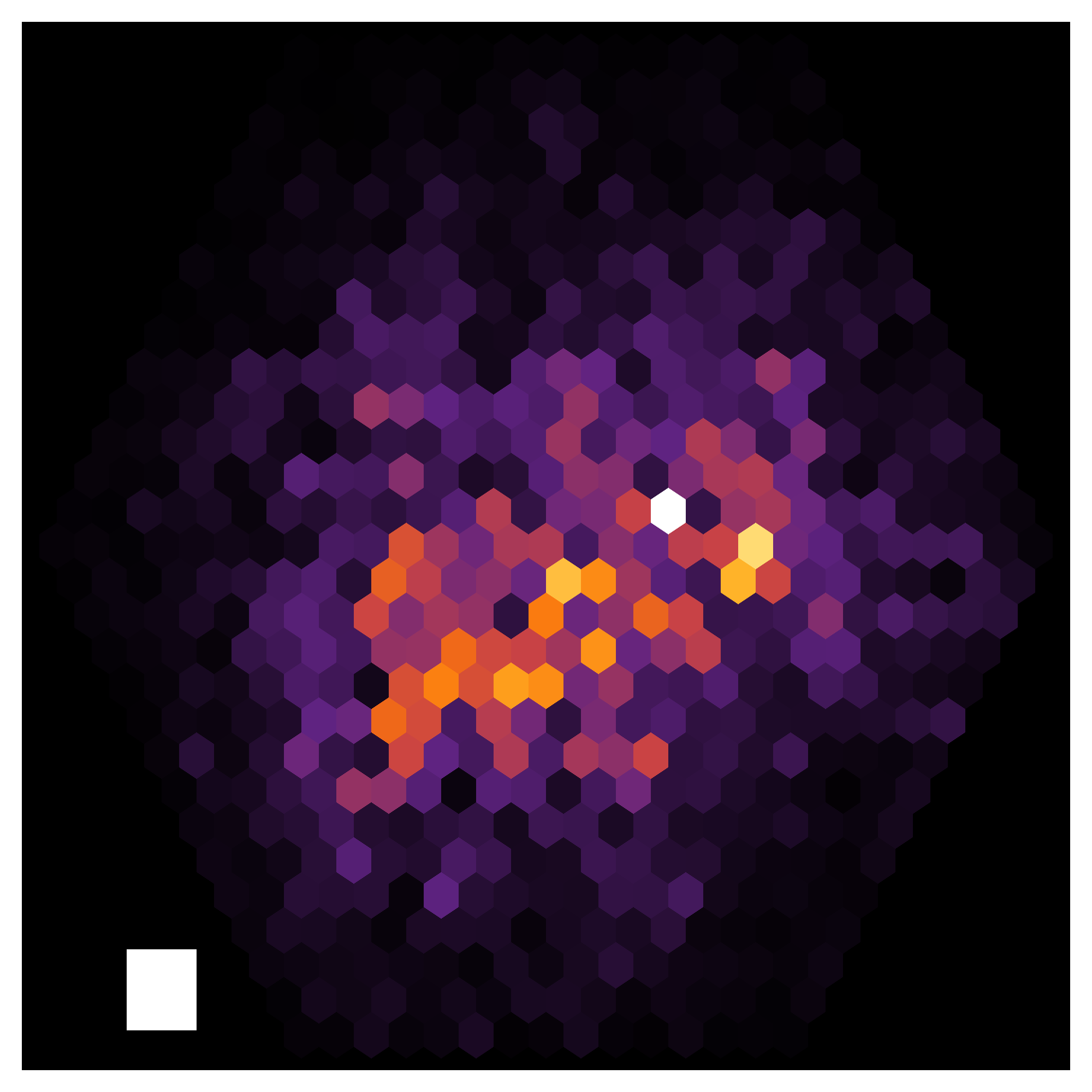}
\includegraphics[width=0.75\columnwidth]{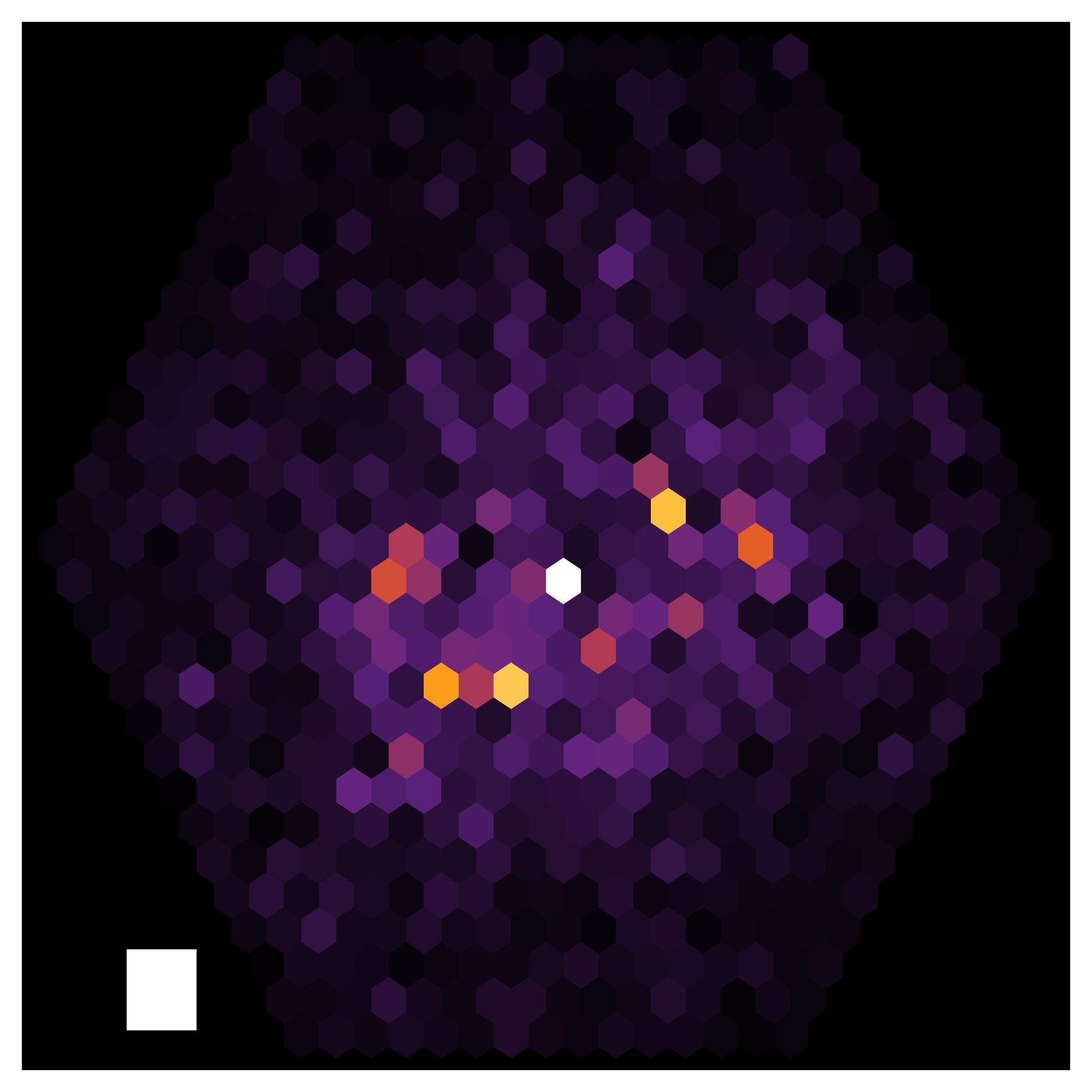}
\includegraphics[width=0.75\columnwidth]{SIFigures/colorbarsunset.png}
\caption{Typical results at system thickness $7.1\, \ell$. Top: Uncorrected image ($C_{\rm uncorr.}=0.521$), Bottom: Corrected image (amplitude and phase correction, $C_{\rm SIM}=0.748$). The white squares indicate the scale of the wavelength in the simulations.}
\label{fig:typicalg25}
\end{figure}

Results for a more strongly scattering layer are  shown in Fig.~\ref{fig:typicalg25}. Now, the uncorrected image does not reveal the object, but the corrected image shows it quite clearly.
The resulting ensemble-averaged correlation between the object and the image formed by an uncorrected (ballistic) focus ($C_{\rm uncorr.}$), by a phase and amplitude correction ($C_{\rm SIM}$) and by a phase-only correction of the SIMs ($C_{\rm SIMpo}$, where the weighting of the SIMs with fidelity has been omitted) is given in Table~\ref{table:statistics}. We observe that the image correction improves the correlation with the true object over a range of optical thickness values. At higher optical thickness (around $L/\ell=10$) the correction method starts to fail because the phases of the eigenvalues become isotropically distributed and the reference phase can no longer be identified. 

\begin{table}[h!]
\centering
\begin{tabularx}{0.49\textwidth} { 
  | >{\raggedright\arraybackslash}X  
  |>{\centering\arraybackslash}c 
  |>{\centering\arraybackslash}c 
  |>{\centering\arraybackslash}c
  | >{\centering\arraybackslash}c | }
 \hline
 $L/\ell$& $C_{\rm uncorr.}$ (sd) & $C_{\rm SIM}$ (sd)   & $C_{\rm SIMpo}$ (sd) & $N$   \\
 \hline
 5.4 & 0.728 (0.03) & 0.802 (0.03) & 0.767 (0.03) & 23\\
\hline
 7.1 & 0.500 (0.07) & 0.606 (0.07) & 0.552 (0.08) & 50\\
\hline
 9.0 & 0.326  (0.07) & 0.395 (0.09) & 0.332  (0.09) & 52   \\
\hline
\end{tabularx}
\caption{Summary of mean correlation coefficients and standard deviations (sd)  for uncorrected fields ($C_{\rm uncorr.}$), for phase- and amplitude-corrected fields ($C_{\rm SIM}$) and for fields where only the phase of the SIM was corrected ($C_{\rm SIMpo}$). The number of independent simulation runs corresponding to different disorder configurations is given as $N$. \label{table:statistics}}
\end{table}

\begin{figure}[h!]
\centering
\includegraphics[width=0.85\columnwidth]{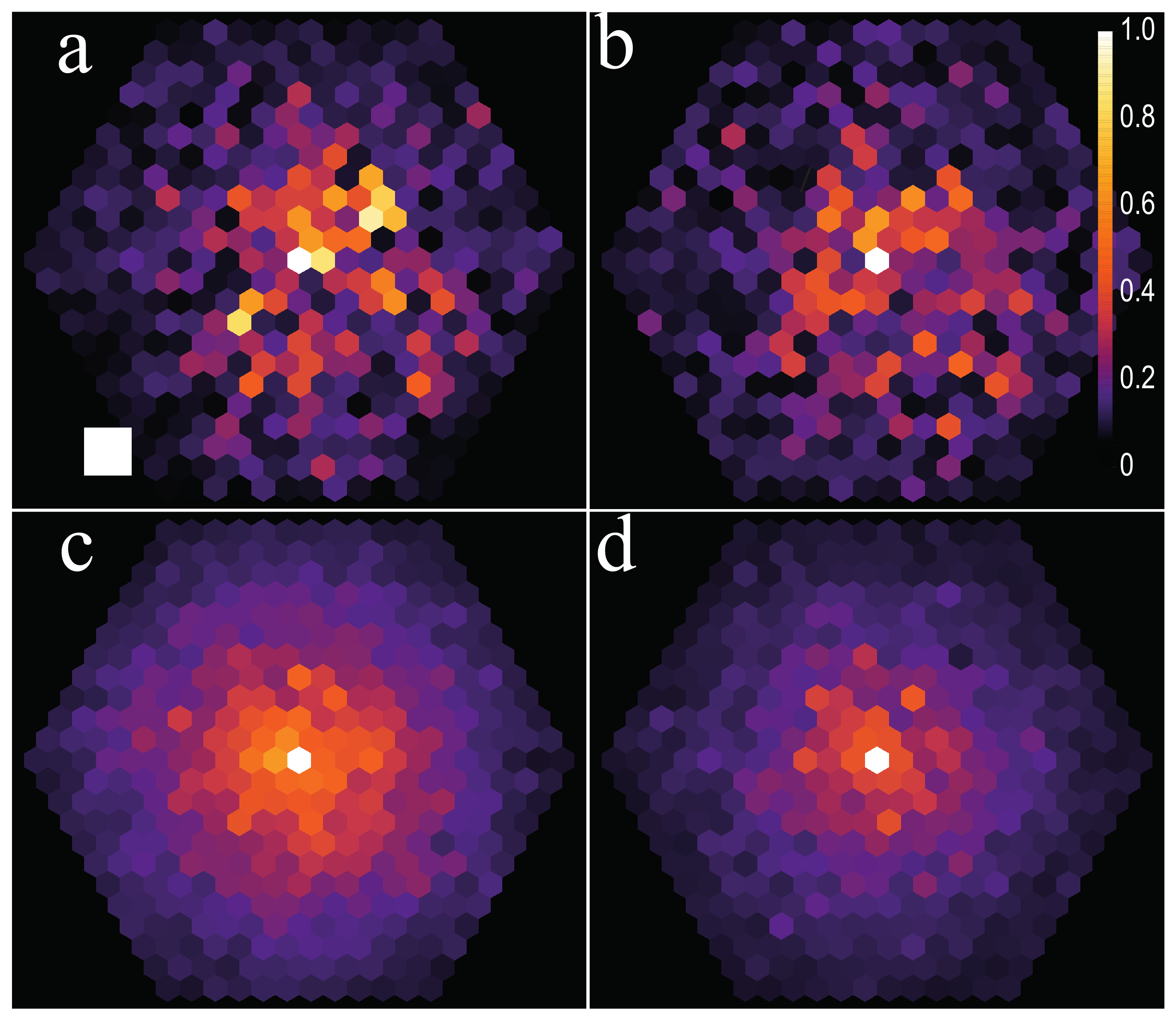}
\caption{Excitation field strength in the center of a sample of thickness $L=7 \ell$. All fields are normalized to the maximum intensity pixel. (a) typical uncorrected field, (b) corresponding SIM-corrected field. (c,d) Root mean square of (a,b) over 50 realizations.}
\label{fig:excitationfields}
\end{figure}

Apart from characterizing the performance on a simple test image, it is useful to consider the shape of the excitation fields before and after correction. In Fig.~\ref{fig:excitationfields} we show a typical example of the field in the object plane in the center of a sample before correction (Fig.~\ref{fig:excitationfields}a) and after SIM correction (Fig.~\ref{fig:excitationfields}b), as well as the corresponding RMS fields averaged over 50 realizations (Fig.~\ref{fig:excitationfields}c,d).
We see that the excitation field is more concentrated in the center pixel after correction. This is also evident from the power fraction in the central pixel (calculated over the visible image), which is on average $5\%$ before, and $9\%$ after correction. Hence we conclude the SIM correction makes the excitation field distribution narrower, while, for the current sample geometry, it almost doubles the power fraction in the focus.

\begin{figure*}[tb]
\centering
\includegraphics[width=0.85\textwidth]{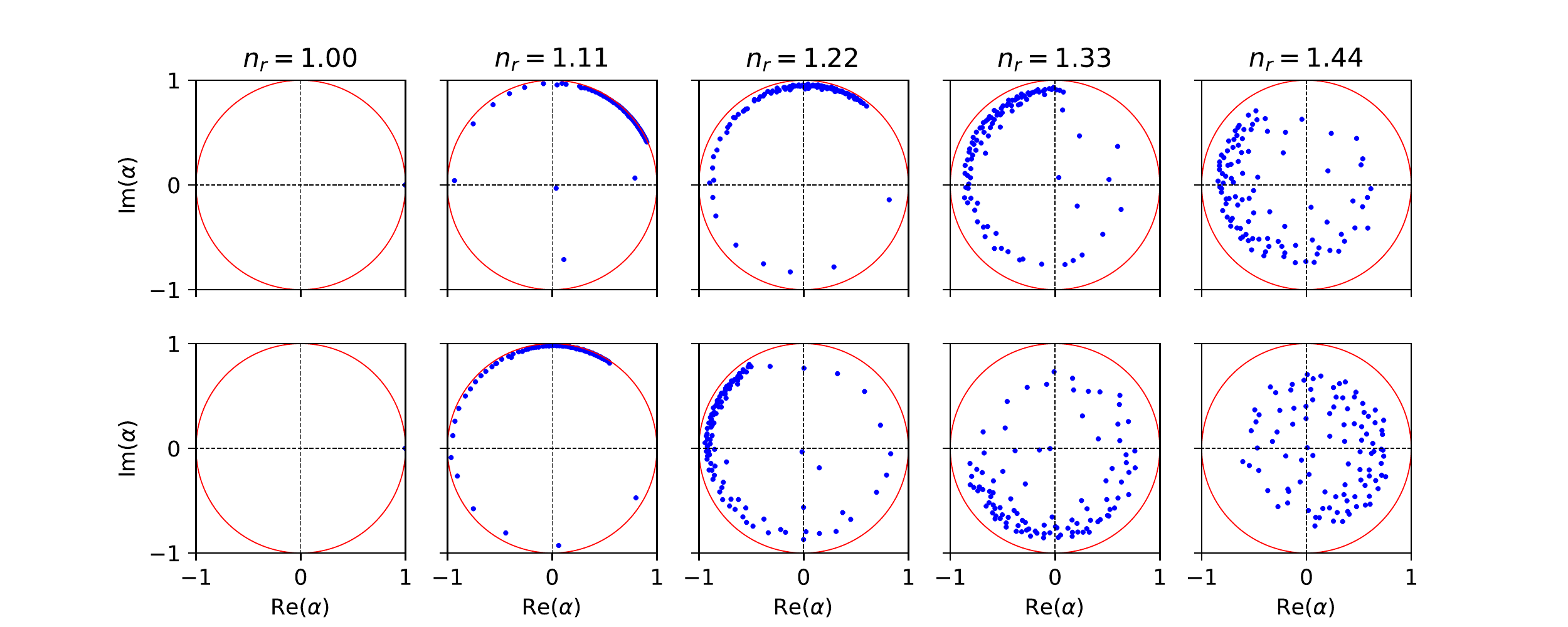}
\caption{Evolution of SIM eigenvalues in the complex plane when turning up the refractive index $n_r$ of the circular scatterers in the two scattering layers in 2D simulations (see supplementary videos for corresponding animations). The two rows correspond to the first two rows in Table~\ref{table:statistics_2D}, where one specific configuration was chosen.}
\label{fig:SIM_eval_evolution}
\end{figure*}
\subsection{Conclusion}
To conclude this section, we have shown that SIMs are a useful basis for implementing methods to estimate and correct the field \textit{inside} a scattering medium. Specifically, we have demonstrated numerically that SIMs can be useful to obtain images in situations where scattering typically makes imaging impractical such as for systems with a thickness in the range between a few and about ten  mean free paths. It may be possible to use SIM-based corrections also in thicker samples by taking into account additional information such as simultaneous retrieval of intensity and amplitude transmission matrices \cite{Boniface2020} or by machine learning algorithms.

We note that in cases where the fluorescent object is bright enough, confocal or rescanned-confocal detection may be used to improve the image quality. This will be true for both the ballistic (uncorrected) method and the method where SIMs are used to correct the excitation beam. The broadband character of typical fluorescence will further help to reduce far-out-of-focus speckle artefacts, which will particularly improve the SIM-corrected result. In the case of weak scattering the SIM-based method becomes a wavefront-sensorless adaptive optics method, which does not need to use fluorescence photons in order to estimate the wavefront correction.
\begin{figure*}[tb]
\centering
\includegraphics[width=0.9\textwidth]{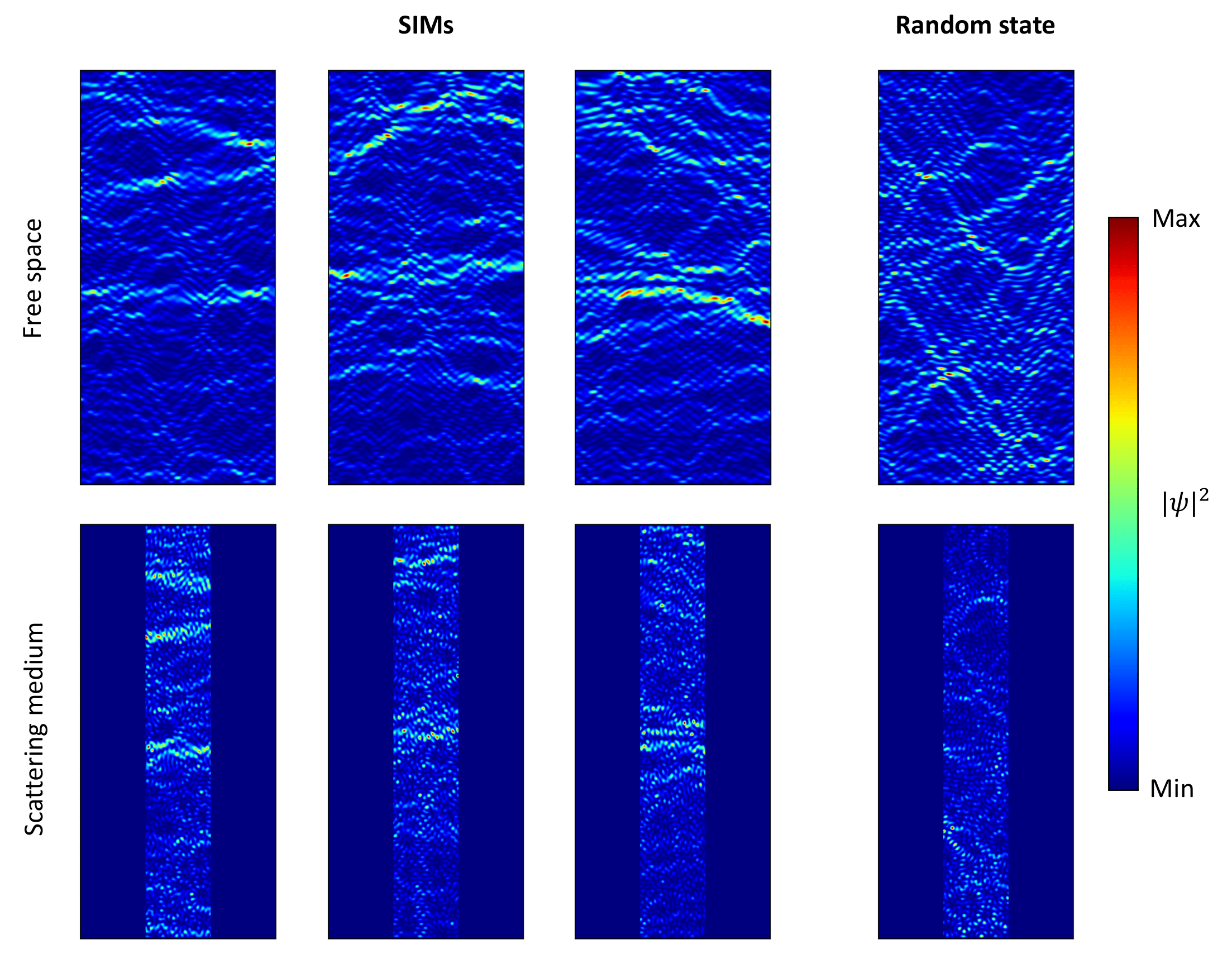}
\caption{Simulated intensity distributions of SIMs with a high $|\alpha|$ (left three columns) and a random state (right column) propagated through air (top row) and through the scattering medium (bottom row). Here, the latter consists of two scattering layers with a thickness of $0.56\, \ell$, where only the propagation in the free space section in the middle is shown.}
\label{fig:SIM_numerical_wavefunctions_correlation}
\end{figure*}
\section{Numerical simulations of imaging internal objects in 2D}
\label{sec:Imaging_2D}
In the case of two-dimensional systems, we employ our full-wave finite-element simulations (see the corresponding section in the methods part of the main text).
We find qualitatively similar results  to  section \ref{sec:Imaging_3D}, however the heuristic formulas  found in the 3D simulation cannot be applied without correction. This is because the reduced dimensionality in 2D lets some of the corresponding SIM eigenvalues revolve by different multiples of $2\pi$ in the complex plane during propagation (see Fig.~\ref{fig:SIM_eval_evolution}). In this way not all of the actual SIM phases can be predicted by using the argument of their eigenvalues only. As mentioned in the previous section, this is caused by the different effective refractive indices of the reference and the principal medium. Thus, using a reference medium with an effective refractive index similar to the one of the principal medium could yield a noticeable improvement of this method as it will reduce the spiralling of SIM eigenvalues in the complex plane. For the remaining part of this section we will use air as the reference medium.

\begin{table}
	\centering
	\begin{tabularx}{0.35\textwidth} { 
			| >{\raggedright\arraybackslash}X  
			|>{\centering\arraybackslash}c 
			| >{\centering\arraybackslash}c | }
		\hline
		$L/\ell$& $C_{\rm uncorr.}$ (sd) & $C_\mathrm{SIM}$ (sd)   \\
		\hline
		1.12 & 0.903 (0.031) & 0.956 (0.018) \\
		\hline
		1.82 & 0.778 (0.069) & 0.859 (0.073) \\
		\hline
		3.14 & 0.740 (0.079) & 0.809 (0.082) \\
		\hline
	\end{tabularx}
	\caption{Summary of mean correlation coefficients and standard deviations (sd) for different scattering configurations in 2D using the modified imaging protocol described in section~\ref{sec:Imaging_2D}. Here, $C_{\rm uncorr.}$ stands for the correlation using uncorrected fields, whereas $C_\mathrm{SIM}$ denotes the image correlation with SIM-corrected focusing states. Each correlation value is an average over 100 different scattering layer configurations and 20 different fluorescence functions, which we choose to consist of three randomly chosen spots of width $\lambda/2$ in the imaging plane. The first two rows correspond to systems with a filling fraction of 5\% and 10\%, whereas the systems in the last row consist of two concatenated layers with 5\% filling fraction in front and after the imaging plane. \label{table:statistics_2D}}
\end{table}

The fast revolving SIM eigenvalues appear to be the reason that heuristic formulas like equation~\eqref{eq:heuristicphase} and \eqref{eq:fidelity} cannot be found in our 2D case. However, rearranging the spectral decomposition of the SIM-matrix
\begin{equation}
T_\mathrm{s} = T_\mathrm{air}(L) A\, \mathrm{diag}\left( \alpha \right)\, A^{-1} \, \end{equation}
we empirically find that approximating the transmission matrix to the imaging plane at $z = L/2$ by the square-root of the total transmission matrix
\begin{equation}
T_\mathrm{s/2} \approx T_\mathrm{air}(L/2) \, A\, \mathrm{diag}\left( \sqrt{\alpha} \right)\, A^{-1}
\label{eq:T_approx_2D}
\end{equation}
and applying $T_\mathrm{s/2}^\dagger$ to $E_{\rm foc}(x,z=L/2)$ results in corrected input focusing fields $E_{\rm in, SIM}$ which yield fluorescence images featuring a higher correlation with the desired images inside the scattering medium than the uncorrected ballistic free-space focusing states in samples with lengths up to 4 transport mean free paths. Note, that in equation~\eqref{eq:T_approx_2D} we set the branch-cut of the complex square root function to the angle at which the SIM eigenvalue distribution has its lowest density in the complex plane. For more strongly scattering systems, we find that the SIM eigenvalue distributions in the complex plane lose their asymmetry. Since the latter indicates the presence of ballistic contributions, this protocol is expected to improve imaging as long as the SIM eigenvalue distributions show a pronounced asymmetry.

Table~\ref{table:statistics_2D} summarizes the findings of our 2D numerics, in which we simulated a waveguide system with two scattering layers. Between these two layers a free space section is embedded, where the imaging plane was chosen to be in the middle of this free space section. The scattering layers consist of small circular obstacles with a diameter of $d_\mathrm{scat} \approx \lambda/6$ and a refractive index of $n_r = 1.44$. To characterize the scattering strengths of our layers, we calculate the transport mean free path $\ell_t$, where $\ell_t \approx \ell$ because the scatterers are in the Rayleigh regime, $d_\mathrm{scat} \ll \lambda$. In contrast to the calculations mentioned in the main text, we use 100 waveguide modes here, i.e., $k = 100.5 \pi/W$ with $W$ being the width of our waveguide. To calculate the pixel values of a fluorescence image, we inject all (uncorrected and corrected) focusing states, propagate them to the imaging plane and calculate the overlap of their intensity profile with a fluorescence function which consists of 3 randomly chosen spots of width $\lambda/2$, where we also use Gaussian window functions around the ideal focus spots in order to simulate the diffusion cones resulting from spatially localized input states (simulating the effect of the pinhole in confocal microscopy). The width of these Gaussian window functions is chosen such that $5 \sigma$ equals the penetration depth into the focusing plane, i.e., half of the total waveguide length. The resulting pixel values are then used to calculate the (Pearson) correlation with the ideal image.

To showcase the correlation of SIMs with their ballistic contribution, i.e., their propagation through air, we plot in Fig.~\ref{fig:SIM_numerical_wavefunctions_correlation} the intensity distributions of SIMs propagated through two scattering layers. Correlations are clearly visible when comparing their intensity profiles to the case of propagation through air (in contrast to the case of random states).

\bibliography{references}

\begin{thebibliography}{41}
\expandafter\ifx\csname natexlab\endcsname\relax\def\natexlab#1{#1}\fi
\expandafter\ifx\csname bibnamefont\endcsname\relax
  \def\bibnamefont#1{#1}\fi
\expandafter\ifx\csname bibfnamefont\endcsname\relax
  \def\bibfnamefont#1{#1}\fi
\expandafter\ifx\csname citenamefont\endcsname\relax
  \def\citenamefont#1{#1}\fi
\expandafter\ifx\csname url\endcsname\relax
  \def\url#1{\texttt{#1}}\fi
\expandafter\ifx\csname urlprefix\endcsname\relax\def\urlprefix{URL }\fi
\providecommand{\bibinfo}[2]{#2}
\providecommand{\eprint}[2][]{\url{#2}}

\bibitem[{\citenamefont{Johnson and Gabriel}(2015)}]{Johnson2015}
\bibinfo{author}{\bibfnamefont{C.}~\bibnamefont{Johnson}} \bibnamefont{and}
  \bibinfo{author}{\bibfnamefont{D.}~\bibnamefont{Gabriel}},
  \emph{\bibinfo{title}{Laser Light Scattering}}, Dover Books on Physics
  (\bibinfo{publisher}{Dover Publications}, \bibinfo{year}{2015}), ISBN
  \bibinfo{isbn}{9780486152202},
  \urlprefix\url{https://books.google.at/books?id=B9-q3LV6xpkC}.

\bibitem[{\citenamefont{Mosk et~al.}(2012)\citenamefont{Mosk, Lagendijk,
  Lerosey, and Fink}}]{Mosk2012}
\bibinfo{author}{\bibfnamefont{A.~P.} \bibnamefont{Mosk}},
  \bibinfo{author}{\bibfnamefont{A.}~\bibnamefont{Lagendijk}},
  \bibinfo{author}{\bibfnamefont{G.}~\bibnamefont{Lerosey}}, \bibnamefont{and}
  \bibinfo{author}{\bibfnamefont{M.}~\bibnamefont{Fink}},
  \bibinfo{journal}{Nature Photonics} \textbf{\bibinfo{volume}{6}},
  \bibinfo{pages}{283} (\bibinfo{year}{2012}), ISSN \bibinfo{issn}{1749-4893},
  \urlprefix\url{https://doi.org/10.1038/nphoton.2012.88}.

\bibitem[{\citenamefont{Rotter and Gigan}(2017)}]{Rotter2017}
\bibinfo{author}{\bibfnamefont{S.}~\bibnamefont{Rotter}} \bibnamefont{and}
  \bibinfo{author}{\bibfnamefont{S.}~\bibnamefont{Gigan}},
  \bibinfo{journal}{Rev. Mod. Phys.} \textbf{\bibinfo{volume}{89}},
  \bibinfo{pages}{015005} (\bibinfo{year}{2017}),
  \urlprefix\url{https://link.aps.org/doi/10.1103/RevModPhys.89.015005}.

\bibitem[{\citenamefont{Yoon et~al.}(2020)\citenamefont{Yoon, Kim, Jang, Choi,
  Choi, Kang, and Choi}}]{Yoon2020}
\bibinfo{author}{\bibfnamefont{S.}~\bibnamefont{Yoon}},
  \bibinfo{author}{\bibfnamefont{M.}~\bibnamefont{Kim}},
  \bibinfo{author}{\bibfnamefont{M.}~\bibnamefont{Jang}},
  \bibinfo{author}{\bibfnamefont{Y.}~\bibnamefont{Choi}},
  \bibinfo{author}{\bibfnamefont{W.}~\bibnamefont{Choi}},
  \bibinfo{author}{\bibfnamefont{S.}~\bibnamefont{Kang}}, \bibnamefont{and}
  \bibinfo{author}{\bibfnamefont{W.}~\bibnamefont{Choi}},
  \bibinfo{journal}{Nature Reviews Physics} \textbf{\bibinfo{volume}{2}},
  \bibinfo{pages}{141} (\bibinfo{year}{2020}), ISSN \bibinfo{issn}{2522-5820},
  \urlprefix\url{https://doi.org/10.1038/s42254-019-0143-2}.

\bibitem[{\citenamefont{Kubby et~al.}(2019)\citenamefont{Kubby, Gigan, and
  Cui}}]{Kubby2020}
\bibinfo{editor}{\bibfnamefont{J.}~\bibnamefont{Kubby}},
  \bibinfo{editor}{\bibfnamefont{S.}~\bibnamefont{Gigan}}, \bibnamefont{and}
  \bibinfo{editor}{\bibfnamefont{M.}~\bibnamefont{Cui}}, eds.,
  \emph{\bibinfo{title}{Wavefront Shaping for Biomedical Imaging}}
  (\bibinfo{publisher}{Cambridge University Press}, \bibinfo{year}{2019}).

\bibitem[{\citenamefont{Bertolotti et~al.}(2012)\citenamefont{Bertolotti, van
  Putten, Blum, Lagendijk, Vos, and Mosk}}]{Bertolotti2012}
\bibinfo{author}{\bibfnamefont{J.}~\bibnamefont{Bertolotti}},
  \bibinfo{author}{\bibfnamefont{E.~G.} \bibnamefont{van Putten}},
  \bibinfo{author}{\bibfnamefont{C.}~\bibnamefont{Blum}},
  \bibinfo{author}{\bibfnamefont{A.}~\bibnamefont{Lagendijk}},
  \bibinfo{author}{\bibfnamefont{W.~L.} \bibnamefont{Vos}}, \bibnamefont{and}
  \bibinfo{author}{\bibfnamefont{A.~P.} \bibnamefont{Mosk}},
  \bibinfo{journal}{Nature} \textbf{\bibinfo{volume}{491}},
  \bibinfo{pages}{232} (\bibinfo{year}{2012}).

\bibitem[{\citenamefont{Katz et~al.}(2014)\citenamefont{Katz, Heidmann, Fink,
  and Gigan}}]{Katz2014_nphot}
\bibinfo{author}{\bibfnamefont{O.}~\bibnamefont{Katz}},
  \bibinfo{author}{\bibfnamefont{P.}~\bibnamefont{Heidmann}},
  \bibinfo{author}{\bibfnamefont{M.}~\bibnamefont{Fink}}, \bibnamefont{and}
  \bibinfo{author}{\bibfnamefont{S.}~\bibnamefont{Gigan}},
  \bibinfo{journal}{Nature Photonics} \textbf{\bibinfo{volume}{8}},
  \bibinfo{pages}{784} (\bibinfo{year}{2014}).

\bibitem[{\citenamefont{Kang et~al.}(2017)\citenamefont{Kang, Kang, Jeong,
  Kwon, Yang, Hong, Kim, Song, Park, Lee et~al.}}]{Kang2017}
\bibinfo{author}{\bibfnamefont{S.}~\bibnamefont{Kang}},
  \bibinfo{author}{\bibfnamefont{P.}~\bibnamefont{Kang}},
  \bibinfo{author}{\bibfnamefont{S.}~\bibnamefont{Jeong}},
  \bibinfo{author}{\bibfnamefont{Y.}~\bibnamefont{Kwon}},
  \bibinfo{author}{\bibfnamefont{T.~D.} \bibnamefont{Yang}},
  \bibinfo{author}{\bibfnamefont{J.~H.} \bibnamefont{Hong}},
  \bibinfo{author}{\bibfnamefont{M.}~\bibnamefont{Kim}},
  \bibinfo{author}{\bibfnamefont{K.}~\bibnamefont{Song}},
  \bibinfo{author}{\bibfnamefont{J.~H.} \bibnamefont{Park}},
  \bibinfo{author}{\bibfnamefont{J.~H.} \bibnamefont{Lee}},
  \bibnamefont{et~al.}, \bibinfo{journal}{Nature Communications}
  \textbf{\bibinfo{volume}{8}} (\bibinfo{year}{2017}).

\bibitem[{\citenamefont{Jang et~al.}(2018)\citenamefont{Jang, Horie, Shibukawa,
  Brake, Liu, Kamali, Arbabi, Ruan, Faraon, and Yang}}]{Jang2018}
\bibinfo{author}{\bibfnamefont{M.}~\bibnamefont{Jang}},
  \bibinfo{author}{\bibfnamefont{Y.}~\bibnamefont{Horie}},
  \bibinfo{author}{\bibfnamefont{A.}~\bibnamefont{Shibukawa}},
  \bibinfo{author}{\bibfnamefont{J.}~\bibnamefont{Brake}},
  \bibinfo{author}{\bibfnamefont{Y.}~\bibnamefont{Liu}},
  \bibinfo{author}{\bibfnamefont{S.~M.} \bibnamefont{Kamali}},
  \bibinfo{author}{\bibfnamefont{A.}~\bibnamefont{Arbabi}},
  \bibinfo{author}{\bibfnamefont{H.}~\bibnamefont{Ruan}},
  \bibinfo{author}{\bibfnamefont{A.}~\bibnamefont{Faraon}}, \bibnamefont{and}
  \bibinfo{author}{\bibfnamefont{C.}~\bibnamefont{Yang}},
  \bibinfo{journal}{Nature Photonics} \textbf{\bibinfo{volume}{12}},
  \bibinfo{pages}{84} (\bibinfo{year}{2018}).

\bibitem[{\citenamefont{Horisaki et~al.}(2019)\citenamefont{Horisaki, Mori, and
  Tanida}}]{Horisaki2019}
\bibinfo{author}{\bibfnamefont{R.}~\bibnamefont{Horisaki}},
  \bibinfo{author}{\bibfnamefont{Y.}~\bibnamefont{Mori}}, \bibnamefont{and}
  \bibinfo{author}{\bibfnamefont{J.}~\bibnamefont{Tanida}},
  \bibinfo{journal}{Optical Review} \textbf{\bibinfo{volume}{26}},
  \bibinfo{pages}{709} (\bibinfo{year}{2019}).

\bibitem[{\citenamefont{Badon et~al.}()\citenamefont{Badon, Barolle, Irsch,
  Boccara, Fink, and Aubry}}]{badon2019}
\bibinfo{author}{\bibfnamefont{A.}~\bibnamefont{Badon}},
  \bibinfo{author}{\bibfnamefont{V.}~\bibnamefont{Barolle}},
  \bibinfo{author}{\bibfnamefont{K.}~\bibnamefont{Irsch}},
  \bibinfo{author}{\bibfnamefont{A.~C.} \bibnamefont{Boccara}},
  \bibinfo{author}{\bibfnamefont{M.}~\bibnamefont{Fink}}, \bibnamefont{and}
  \bibinfo{author}{\bibfnamefont{A.}~\bibnamefont{Aubry}},
  \emph{\bibinfo{title}{Distortion matrix concept for deep imaging in optical
  coherence microscopy.}}, \bibinfo{note}{preprint at
  \href{https://arxiv.org/abs/1910.07252}{https://arxiv.org/abs/1910.07252}
  (2019)}, \eprint{1910.07252}.

\bibitem[{\citenamefont{Wang and Wu}(2007)}]{Wang2007}
\bibinfo{author}{\bibfnamefont{L.~V.} \bibnamefont{Wang}} \bibnamefont{and}
  \bibinfo{author}{\bibfnamefont{H.}~\bibnamefont{Wu}},
  \emph{\bibinfo{title}{Biomedical Optics: Principles and Imaging}}
  (\bibinfo{publisher}{Wiley-Interscience}, \bibinfo{year}{2007}).

\bibitem[{\citenamefont{Ntziachristos}(2010)}]{Ntziachristos2010}
\bibinfo{author}{\bibfnamefont{V.}~\bibnamefont{Ntziachristos}},
  \bibinfo{journal}{Nature Methods} \textbf{\bibinfo{volume}{7}},
  \bibinfo{pages}{603} (\bibinfo{year}{2010}), ISSN \bibinfo{issn}{1548-7105}.

\bibitem[{\citenamefont{Drexler and Fujimoto}(2015)}]{Drexler2015}
\bibinfo{editor}{\bibfnamefont{W.}~\bibnamefont{Drexler}} \bibnamefont{and}
  \bibinfo{editor}{\bibfnamefont{J.}~\bibnamefont{Fujimoto}}, eds.,
  \emph{\bibinfo{title}{Optical Coherence Tomography: Technology and
  Applications}} (\bibinfo{publisher}{Springer International Publishing},
  \bibinfo{year}{2015}).

\bibitem[{\citenamefont{Dorokhov}(1984)}]{Dorokhov1984}
\bibinfo{author}{\bibfnamefont{O.}~\bibnamefont{Dorokhov}},
  \bibinfo{journal}{Solid State Communications} \textbf{\bibinfo{volume}{51}},
  \bibinfo{pages}{381 } (\bibinfo{year}{1984}), ISSN \bibinfo{issn}{0038-1098},
  \urlprefix\url{http://www.sciencedirect.com/science/article/pii/0038109884901170}.

\bibitem[{\citenamefont{Beenakker}(1997)}]{beenakker}
\bibinfo{author}{\bibfnamefont{C.~W.~J.} \bibnamefont{Beenakker}},
  \bibinfo{journal}{Rev. Mod. Phys.} \textbf{\bibinfo{volume}{69}},
  \bibinfo{pages}{731} (\bibinfo{year}{1997}),
  \urlprefix\url{https://link.aps.org/doi/10.1103/RevModPhys.69.731}.

\bibitem[{\citenamefont{Pendry et~al.}(1990)\citenamefont{Pendry, MacKinnon,
  and Pretre}}]{pendry1990}
\bibinfo{author}{\bibfnamefont{J.}~\bibnamefont{Pendry}},
  \bibinfo{author}{\bibfnamefont{A.}~\bibnamefont{MacKinnon}},
  \bibnamefont{and} \bibinfo{author}{\bibfnamefont{A.}~\bibnamefont{Pretre}},
  \bibinfo{journal}{Physica A: Statistical Mechanics and its Applications}
  \textbf{\bibinfo{volume}{168}}, \bibinfo{pages}{400} (\bibinfo{year}{1990}).

\bibitem[{\citenamefont{Akkermans and
  Montambaux}(2007)}]{akkermans_montambaux_2007}
\bibinfo{author}{\bibfnamefont{E.}~\bibnamefont{Akkermans}} \bibnamefont{and}
  \bibinfo{author}{\bibfnamefont{G.}~\bibnamefont{Montambaux}},
  \emph{\bibinfo{title}{Mesoscopic Physics of Electrons and Photons}}
  (\bibinfo{publisher}{Cambridge University Press}, \bibinfo{year}{2007}).

\bibitem[{\citenamefont{Peña et~al.}(2014)\citenamefont{Peña, Girschik,
  Libisch, Rotter, and Chabanov}}]{pena2014}
\bibinfo{author}{\bibfnamefont{A.}~\bibnamefont{Peña}},
  \bibinfo{author}{\bibfnamefont{A.}~\bibnamefont{Girschik}},
  \bibinfo{author}{\bibfnamefont{F.}~\bibnamefont{Libisch}},
  \bibinfo{author}{\bibfnamefont{S.}~\bibnamefont{Rotter}}, \bibnamefont{and}
  \bibinfo{author}{\bibfnamefont{A.~A.} \bibnamefont{Chabanov}},
  \bibinfo{journal}{Nature Communications} \textbf{\bibinfo{volume}{5}},
  \bibinfo{pages}{3488} (\bibinfo{year}{2014}), ISSN \bibinfo{issn}{2041-1723},
  \urlprefix\url{https://doi.org/10.1038/ncomms4488}.

\bibitem[{\citenamefont{Davy et~al.}(2013)\citenamefont{Davy, Shi, Wang, and
  Genack}}]{Davy13}
\bibinfo{author}{\bibfnamefont{M.}~\bibnamefont{Davy}},
  \bibinfo{author}{\bibfnamefont{Z.}~\bibnamefont{Shi}},
  \bibinfo{author}{\bibfnamefont{J.}~\bibnamefont{Wang}}, \bibnamefont{and}
  \bibinfo{author}{\bibfnamefont{A.~Z.} \bibnamefont{Genack}},
  \bibinfo{journal}{Opt. Express} \textbf{\bibinfo{volume}{21}},
  \bibinfo{pages}{10367} (\bibinfo{year}{2013}),
  \urlprefix\url{http://www.opticsexpress.org/abstract.cfm?URI=oe-21-8-10367}.

\bibitem[{\citenamefont{{Miller}}(2019)}]{miller2019}
\bibinfo{author}{\bibfnamefont{D.~A.~B.} \bibnamefont{{Miller}}},
  \bibinfo{journal}{arXiv e-prints} \bibinfo{eid}{arXiv:1904.05427}
  (\bibinfo{year}{2019}), \eprint{1904.05427}.

\bibitem[{\citenamefont{Vellekoop and Mosk}(2008)}]{vellekoop2008prl}
\bibinfo{author}{\bibfnamefont{I.~M.} \bibnamefont{Vellekoop}}
  \bibnamefont{and} \bibinfo{author}{\bibfnamefont{A.~P.} \bibnamefont{Mosk}},
  \bibinfo{journal}{Phys. Rev. Lett.} \textbf{\bibinfo{volume}{101}},
  \bibinfo{pages}{120601} (\bibinfo{year}{2008}),
  \urlprefix\url{https://link.aps.org/doi/10.1103/PhysRevLett.101.120601}.

\bibitem[{\citenamefont{Hsu et~al.}(2017)\citenamefont{Hsu, Liew, Goetschy,
  Cao, and Stone}}]{Hsu2017}
\bibinfo{author}{\bibfnamefont{C.~W.} \bibnamefont{Hsu}},
  \bibinfo{author}{\bibfnamefont{S.~F.} \bibnamefont{Liew}},
  \bibinfo{author}{\bibfnamefont{A.}~\bibnamefont{Goetschy}},
  \bibinfo{author}{\bibfnamefont{H.}~\bibnamefont{Cao}}, \bibnamefont{and}
  \bibinfo{author}{\bibfnamefont{A.~D.} \bibnamefont{Stone}},
  \bibinfo{journal}{Nature Physics} \textbf{\bibinfo{volume}{13}},
  \bibinfo{pages}{497} (\bibinfo{year}{2017}).

\bibitem[{\citenamefont{Yu et~al.}(2013)\citenamefont{Yu, Hillman, Choi, Lee,
  Feld, Dasari, and Park}}]{yu2013}
\bibinfo{author}{\bibfnamefont{H.}~\bibnamefont{Yu}},
  \bibinfo{author}{\bibfnamefont{T.~R.} \bibnamefont{Hillman}},
  \bibinfo{author}{\bibfnamefont{W.}~\bibnamefont{Choi}},
  \bibinfo{author}{\bibfnamefont{J.~O.} \bibnamefont{Lee}},
  \bibinfo{author}{\bibfnamefont{M.~S.} \bibnamefont{Feld}},
  \bibinfo{author}{\bibfnamefont{R.~R.} \bibnamefont{Dasari}},
  \bibnamefont{and} \bibinfo{author}{\bibfnamefont{Y.}~\bibnamefont{Park}},
  \bibinfo{journal}{Phys. Rev. Lett.} \textbf{\bibinfo{volume}{111}},
  \bibinfo{pages}{153902} (\bibinfo{year}{2013}),
  \urlprefix\url{https://link.aps.org/doi/10.1103/PhysRevLett.111.153902}.

\bibitem[{\citenamefont{Kim et~al.}(2012)\citenamefont{Kim, Choi, Yoon, Choi,
  Kim, Park, and Choi}}]{kim2012}
\bibinfo{author}{\bibfnamefont{M.}~\bibnamefont{Kim}},
  \bibinfo{author}{\bibfnamefont{Y.}~\bibnamefont{Choi}},
  \bibinfo{author}{\bibfnamefont{C.}~\bibnamefont{Yoon}},
  \bibinfo{author}{\bibfnamefont{W.}~\bibnamefont{Choi}},
  \bibinfo{author}{\bibfnamefont{J.}~\bibnamefont{Kim}},
  \bibinfo{author}{\bibfnamefont{Q.-H.} \bibnamefont{Park}}, \bibnamefont{and}
  \bibinfo{author}{\bibfnamefont{W.}~\bibnamefont{Choi}},
  \bibinfo{journal}{Nature Photonics} \textbf{\bibinfo{volume}{6}},
  \bibinfo{pages}{581} (\bibinfo{year}{2012}), ISSN \bibinfo{issn}{1749-4893},
  \urlprefix\url{https://doi.org/10.1038/nphoton.2012.159}.

\bibitem[{\citenamefont{Mello et~al.}(1988)\citenamefont{Mello, Pereyra, and
  Kumar}}]{DMPK}
\bibinfo{author}{\bibfnamefont{P.}~\bibnamefont{Mello}},
  \bibinfo{author}{\bibfnamefont{P.}~\bibnamefont{Pereyra}}, \bibnamefont{and}
  \bibinfo{author}{\bibfnamefont{N.}~\bibnamefont{Kumar}},
  \bibinfo{journal}{Annals of Physics} \textbf{\bibinfo{volume}{181}},
  \bibinfo{pages}{290 } (\bibinfo{year}{1988}), ISSN \bibinfo{issn}{0003-4916},
  \urlprefix\url{http://www.sciencedirect.com/science/article/pii/0003491688901698}.

\bibitem[{\citenamefont{Bosch et~al.}(2016)\citenamefont{Bosch, Goorden, and
  Mosk}}]{Bosch2016}
\bibinfo{author}{\bibfnamefont{J.}~\bibnamefont{Bosch}},
  \bibinfo{author}{\bibfnamefont{S.~A.} \bibnamefont{Goorden}},
  \bibnamefont{and} \bibinfo{author}{\bibfnamefont{A.~P.} \bibnamefont{Mosk}},
  \bibinfo{journal}{Optics Express} \textbf{\bibinfo{volume}{24}},
  \bibinfo{pages}{26472} (\bibinfo{year}{2016}).

\bibitem[{\citenamefont{Pai et~al.}(2020)\citenamefont{Pai, Bosch, and
  Mosk}}]{Pai20}
\bibinfo{author}{\bibfnamefont{P.}~\bibnamefont{Pai}},
  \bibinfo{author}{\bibfnamefont{J.}~\bibnamefont{Bosch}}, \bibnamefont{and}
  \bibinfo{author}{\bibfnamefont{A.~P.} \bibnamefont{Mosk}},
  \bibinfo{journal}{OSA Continuum} \textbf{\bibinfo{volume}{3}},
  \bibinfo{pages}{637} (\bibinfo{year}{2020}),
  \urlprefix\url{http://www.osapublishing.org/osac/abstract.cfm?URI=osac-3-3-637}.

\bibitem[{\citenamefont{Goetschy and Stone}(2013)}]{stone2013}
\bibinfo{author}{\bibfnamefont{A.}~\bibnamefont{Goetschy}} \bibnamefont{and}
  \bibinfo{author}{\bibfnamefont{A.~D.} \bibnamefont{Stone}},
  \bibinfo{journal}{Phys. Rev. Lett.} \textbf{\bibinfo{volume}{111}},
  \bibinfo{pages}{063901} (\bibinfo{year}{2013}),
  \urlprefix\url{https://link.aps.org/doi/10.1103/PhysRevLett.111.063901}.

\bibitem[{\citenamefont{Forrester and Nagao}(2007)}]{Forrester2007_prl}
\bibinfo{author}{\bibfnamefont{P.~J.} \bibnamefont{Forrester}}
  \bibnamefont{and} \bibinfo{author}{\bibfnamefont{T.}~\bibnamefont{Nagao}},
  \bibinfo{journal}{Phys. Rev. Lett.} \textbf{\bibinfo{volume}{99}},
  \bibinfo{pages}{050603} (\bibinfo{year}{2007}),
  \urlprefix\url{https://link.aps.org/doi/10.1103/PhysRevLett.99.050603}.

\bibitem[{\citenamefont{Popoff et~al.}(2010)\citenamefont{Popoff, Lerosey,
  Carminati, Fink, Boccara, and Gigan}}]{popoff2010}
\bibinfo{author}{\bibfnamefont{S.~M.} \bibnamefont{Popoff}},
  \bibinfo{author}{\bibfnamefont{G.}~\bibnamefont{Lerosey}},
  \bibinfo{author}{\bibfnamefont{R.}~\bibnamefont{Carminati}},
  \bibinfo{author}{\bibfnamefont{M.}~\bibnamefont{Fink}},
  \bibinfo{author}{\bibfnamefont{A.~C.} \bibnamefont{Boccara}},
  \bibnamefont{and} \bibinfo{author}{\bibfnamefont{S.}~\bibnamefont{Gigan}},
  \bibinfo{journal}{Phys. Rev. Lett.} \textbf{\bibinfo{volume}{104}},
  \bibinfo{pages}{100601} (\bibinfo{year}{2010}),
  \urlprefix\url{https://link.aps.org/doi/10.1103/PhysRevLett.104.100601}.

\bibitem[{\citenamefont{Cuche et~al.}(2000)\citenamefont{Cuche, Marquet, and
  Depeursinge}}]{Cuche00}
\bibinfo{author}{\bibfnamefont{E.}~\bibnamefont{Cuche}},
  \bibinfo{author}{\bibfnamefont{P.}~\bibnamefont{Marquet}}, \bibnamefont{and}
  \bibinfo{author}{\bibfnamefont{C.}~\bibnamefont{Depeursinge}},
  \bibinfo{journal}{Appl. Opt.} \textbf{\bibinfo{volume}{39}},
  \bibinfo{pages}{4070} (\bibinfo{year}{2000}),
  \urlprefix\url{http://ao.osa.org/abstract.cfm?URI=ao-39-23-4070}.

\bibitem[{\citenamefont{Sch{\"o}berl}()}]{Schoeberl2014}
\bibinfo{author}{\bibfnamefont{J.}~\bibnamefont{Sch{\"o}berl}},
  \emph{\bibinfo{title}{{C++11 Implementation of Finite Elements in NGSolve}}},
  \bibinfo{howpublished}{ASC Report, Institute for Analysis and Scientific
  Computing, Vienna University of Technology (2014)}.

\bibitem[{\citenamefont{Lee}(1974)}]{Lee1974}
\bibinfo{author}{\bibfnamefont{W.-H.} \bibnamefont{Lee}},
  \bibinfo{journal}{Applied Optics} \textbf{\bibinfo{volume}{13}},
  \bibinfo{pages}{1677} (\bibinfo{year}{1974}).

\bibitem[{\citenamefont{Leith and Upatnieks}(1962)}]{Leith62}
\bibinfo{author}{\bibfnamefont{E.~N.} \bibnamefont{Leith}} \bibnamefont{and}
  \bibinfo{author}{\bibfnamefont{J.}~\bibnamefont{Upatnieks}},
  \bibinfo{journal}{J. Opt. Soc. Am.} \textbf{\bibinfo{volume}{52}},
  \bibinfo{pages}{1123} (\bibinfo{year}{1962}),
  \urlprefix\url{http://www.osapublishing.org/abstract.cfm?URI=josa-52-10-1123}.

\bibitem[{\citenamefont{Takeda et~al.}(1982)\citenamefont{Takeda, Ina, and
  Kobayashi}}]{Takeda82}
\bibinfo{author}{\bibfnamefont{M.}~\bibnamefont{Takeda}},
  \bibinfo{author}{\bibfnamefont{H.}~\bibnamefont{Ina}}, \bibnamefont{and}
  \bibinfo{author}{\bibfnamefont{S.}~\bibnamefont{Kobayashi}},
  \bibinfo{journal}{J. Opt. Soc. Am.} \textbf{\bibinfo{volume}{72}},
  \bibinfo{pages}{156} (\bibinfo{year}{1982}),
  \urlprefix\url{http://www.osapublishing.org/abstract.cfm?URI=josa-72-1-156}.

\bibitem[{\citenamefont{Banon et~al.}(2020)\citenamefont{Banon, Simonsen, and
  Carminati}}]{banon2020depolarization}
\bibinfo{author}{\bibfnamefont{J.-P.} \bibnamefont{Banon}},
  \bibinfo{author}{\bibfnamefont{I.}~\bibnamefont{Simonsen}}, \bibnamefont{and}
  \bibinfo{author}{\bibfnamefont{R.}~\bibnamefont{Carminati}},
  \bibinfo{journal}{Phys. Rev. A} \textbf{\bibinfo{volume}{101}},
  \bibinfo{pages}{053847} (\bibinfo{year}{2020}),
  \urlprefix\url{https://link.aps.org/doi/10.1103/PhysRevA.101.053847}.

\bibitem[{\citenamefont{Pai et~al.}(2021)\citenamefont{Pai, Bosch, and
  Mosk}}]{Pai21}
\bibinfo{author}{\bibfnamefont{P.}~\bibnamefont{Pai}},
  \bibinfo{author}{\bibfnamefont{J.}~\bibnamefont{Bosch}}, \bibnamefont{and}
  \bibinfo{author}{\bibfnamefont{A.~P.} \bibnamefont{Mosk}},
  \bibinfo{journal}{Opt. Express} \textbf{\bibinfo{volume}{29}},
  \bibinfo{pages}{24} (\bibinfo{year}{2021}),
  \urlprefix\url{http://www.opticsexpress.org/abstract.cfm?URI=oe-29-1-24}.

\bibitem[{\citenamefont{Kantorovich and Krylov}(1958)}]{kantorovich}
\bibinfo{author}{\bibfnamefont{L.~V.} \bibnamefont{Kantorovich}}
  \bibnamefont{and} \bibinfo{author}{\bibfnamefont{V.~I.}
  \bibnamefont{Krylov}}, \emph{\bibinfo{title}{Approximate methods of higher
  analysis}} (\bibinfo{publisher}{Interscience Publishers},
  \bibinfo{year}{1958}).

\bibitem[{\citenamefont{Ko and Inkson}(1988)}]{Ko1988}
\bibinfo{author}{\bibfnamefont{D.~Y.~K.} \bibnamefont{Ko}} \bibnamefont{and}
  \bibinfo{author}{\bibfnamefont{J.~C.} \bibnamefont{Inkson}},
  \bibinfo{journal}{Phys. Rev. B} \textbf{\bibinfo{volume}{38}},
  \bibinfo{pages}{9945} (\bibinfo{year}{1988}),
  \urlprefix\url{https://link.aps.org/doi/10.1103/PhysRevB.38.9945}.

\bibitem[{\citenamefont{Boniface et~al.}(2020)\citenamefont{Boniface, Dong, and
  Gigan}}]{Boniface2020}
\bibinfo{author}{\bibfnamefont{A.}~\bibnamefont{Boniface}},
  \bibinfo{author}{\bibfnamefont{J.}~\bibnamefont{Dong}}, \bibnamefont{and}
  \bibinfo{author}{\bibfnamefont{S.}~\bibnamefont{Gigan}},
  \bibinfo{journal}{Nature Communications} \textbf{\bibinfo{volume}{11}},
  \bibinfo{pages}{6154} (\bibinfo{year}{2020}), ISSN \bibinfo{issn}{2041-1723},
  \urlprefix\url{https://doi.org/10.1038/s41467-020-19696-8}.

\end{thebibliography}

\end{document}